\documentclass[11pt]{amsart}

\usepackage{float}
\usepackage{amsaddr}
\usepackage{geometry}                
\usepackage{graphicx}
\usepackage{amssymb}
\usepackage{epstopdf}
\usepackage{rotating}
\usepackage[authoryear]{natbib}  
\usepackage[footnotesize]{caption}
\usepackage{algorithm}
\usepackage[noend]{algpseudocode}

\usepackage{footnote}
\makesavenoteenv{tabular}
\makesavenoteenv{table}
\newcommand{\bm}[1]{\mbox{\boldmath $#1$}}
\newcommand{\be}{\begin{equation}}
\newcommand{\ee}{\end{equation}}
\usepackage{mathtools}
\newcommand\iid{\stackrel{\mathclap{\normalfont\mbox{\tiny i.i.d.}}}{\sim}}

\title[A Social Network analysis of the ORIE faculty hiring network]{A Social Network Analysis of the ORIE faculty hiring Network}
\author{E. del Castillo, A. Meyers, and P. Chen}
\address{Engineering Statistics and Machine Learning Laboratory}
\address{Dept. of Industrial \& Manufacturing Engineering}
\address{The Pennsylvania State University, University Park, PA 16802}
\date{March 7, 2018}                                           

\begin{document}
\baselineskip=19pt                  

\maketitle
\begin{abstract}
We study the U.S. Operations Research/Industrial-Systems Engineering (ORIE) faculty hiring network, consisting of 1,179 faculty origin and destination data together with attribute data from 83 ORIE departments. 
A social network analysis of faculty hires can reveal important patterns in an academic field, such as the existence of a hierarchy or sociological aspects such as the presence of communities of departments.
We first statistically test for the existence of a linear hierarchy in the network and for its steepness. We find a near linear hierarchical order of the departments, proposing a new index for hiring networks, which we contrast with other indicators of hierarchy, including published rankings. A single index is not capable to capture the full structure of a complex network, however, so we next fit a latent exponential random graph model (ERGM) to the network, which is able to reproduce its main observed characteristics:  high incidence of self-hiring, skewed out-degree distribution, low density and clustering. Finally, we use the latent variables in the ERGM to simplify the network to one where faculty hires take place among three groups of departments. 
We contrast our findings with those reported for other related disciplines, Computer Science and Business.
\end{abstract}


\newpage

\section{Introduction}
\label{Sec:1}
A faculty hiring network is represented by a graph $G=(V,E)$  composed of vertices $v_i \in V, i=1,...,n$ denoting university departments in a given academic discipline, and directed arcs $e_{ij} \in E$, $i,j = 1,2,...,n$ whose integer value attribute denotes the number of faculty hired by department $v_j$ who received their Ph.D.'s in department $v_i$. A department hiring a Ph.D. from another department generates a directed edge in the network going from the sender department to the receiver department. In the pre-internet era, the study of faculty hiring networks was confined to departments of Sociology \citep{Shichor,Burris2004} where directories with the necessary faculty information existed in print. Today, with most departments and faculty posting their information in personal web pages, studying a faculty hiring network has become easier to do even if no directories are available, provided the faculty information is gathered. There now exist analyses of faculty hiring networks in a broad range of disciplines in the US, such as Political Science \citep{Fowler,SchmidtChingos}, Mathematics \citep{Myers}, Communication \citep{LiuGonzalez}, Business \citep{Clauset}, Computer Science \citep{Clauset,Huang}, Law \citep{Katz}, and History \citep{Clauset}.

Researchers have studied  the hiring and placement patterns of their academic fields for a variety of reasons. One concept these studies attempt to clarify is the {\em prestige} of a department, a rather elusive and inadequately theorized concept \citep{Burris2004} but popular among university administrators. Departmental prestige can be studied as an effect due to the position of the department in a hierarchy existing in the faculty hiring network, as these positions provide a ranking within a discipline. Some authors have considered hiring networks for the purpose of determining the inequality in the production of Ph.Ds \citep{Clauset}, an apparently general phenomena in which a small number of departments produce a large fraction of the Ph.Ds hired as professors. Others have studied these networks to determine inequalities in the hiring process (e.g., of women, see \citet{Way}), or to study the sociological aspects of a discipline \citep{Fowler,Katz}, e.g., whether communities or a dominance hierarchy exists among departments. Hiring and placing of Ph.D. students is also studied in some areas to understand how new ideas are disseminated through a profession. 

In this paper we make available and model the U.S. ORIE hiring faculty network. The network is composed of departments/programs/schools (from now on, "departments") within these 3 inter-related fields. Appendix 1 gives a list of all the 83 ORIE departments considered and how they were selected for this study. The ORIE dataset was compiled in summer 2016. We first perform statistical tests on the existence and steepness of a linear dominance hierarchy among the 83 ORIE departments. Having found strong evidence of such hierarchy, we then introduce the notion of a minimum violation and strength (MVS) measure of departmental prestige, and use it to compute a near linear hierarchy for the ORIE network, contrasting it with other more common measures of individual importance in a social network, including published ORIE rankings (US News and NRC). Given that single measures of department prestige provide an inherently incomplete description of a complex network, we further model the positions of each department in the network as latent variables in an exponential random graph model that allows us to consider the uncertainty in the edge data and permits to find groups of related departments. With these positions, we reduce the ORIE network to a simpler network of faculty hires between and within 3 groups of departments. Throughout this study, we compare the ORIE faculty hiring network with similar networks in related academic disciplines. We conclude with a summary and discussion of the implications of our findings.

\section{Descriptive statistics of the ORIE network compared to those from related disciplines.}
\label{Sec:2}

\begin{figure}
\begin{center}
\resizebox{14cm}{5cm}{\includegraphics{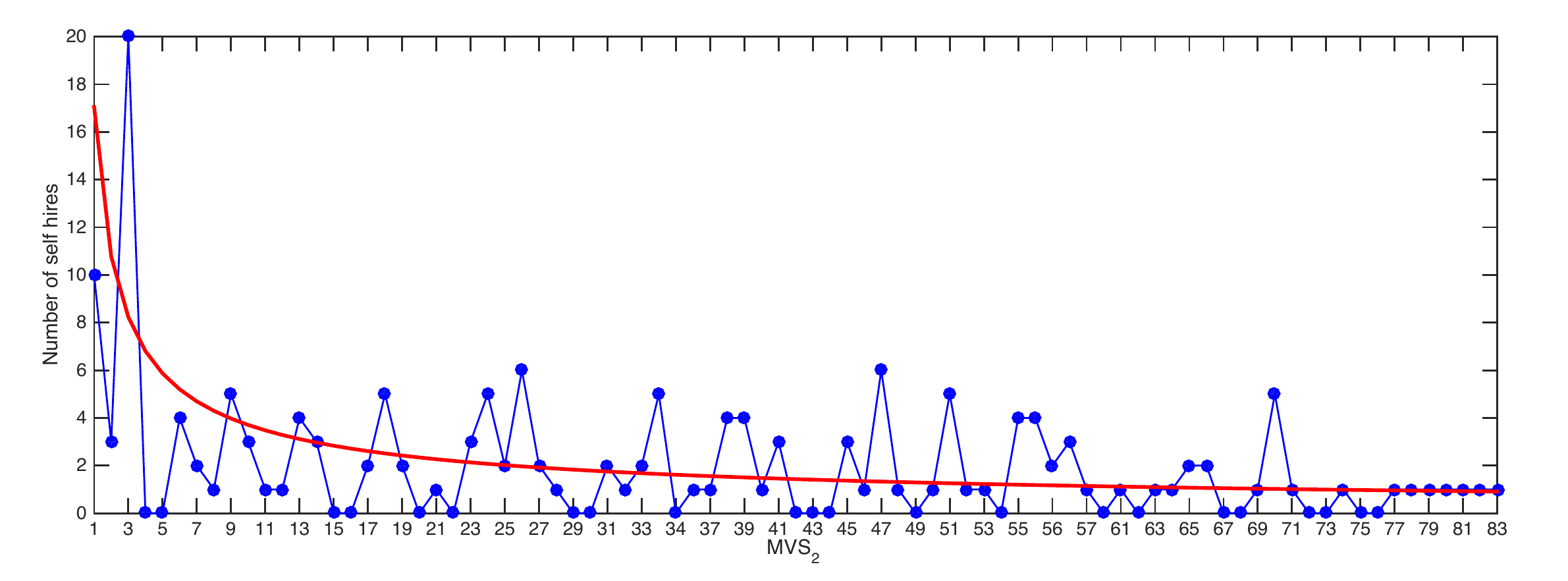}}
\resizebox{13cm}{5cm}{\includegraphics{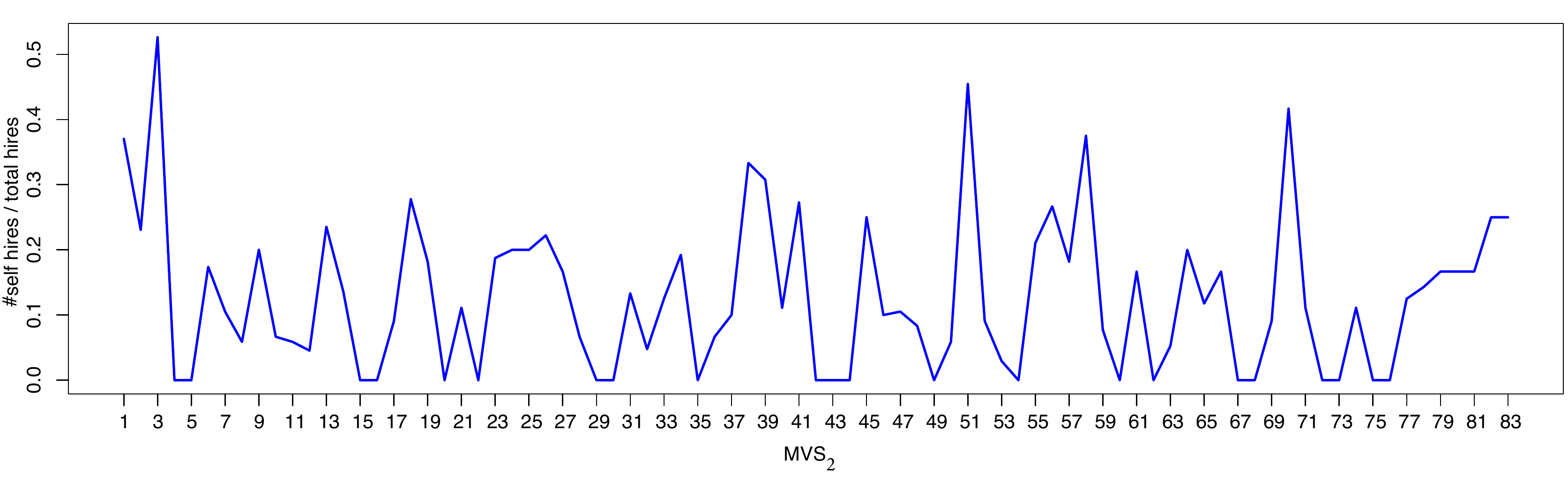}}
\caption{Top: in blue, observed number of self-hires among the 83 ORIE departments sorted by MVS$_2$ hierarchy index described later in this paper.  Bottom: self hires as a percentage of all faculty in each department, sorted by MVS$_2$ index. While self-hires are more numerous in the top departments, self-hires as a proportion of total faculty are significant across all the ORIE departments. In red, expected number of self-hires ($\mu_{ii}$) as predicted by the latent location ERGM model described in section \ref{Sec:5}.}
\label{fig:0} 
\end{center}
\end{figure}

ORIE faculty data were collected during May-June of 2016 from faculty web pages working in institutions in the USA. A list of all Ph.D. granting institutions considered and the criteria used for their inclusion are given in Appendix 1. In total, 1179 faculty from 83 ORIE departments were considered in the analysis. Table \ref{tab:1} lists general descriptive statistics of the ORIE network compared to those of two closely related fields,  Computer Science and Business schools, networks analyzed  by \citet{Clauset}, excluding their ``Earth" (outside US) department and links from and to this vertex. The ORIE network is considerably smaller, with an average department size of 14.2 faculty compared to 21.4 faculty for CS and the much bigger schools of Business with an average number of 70.1 faculty. 
\begin{small}
\begin{table}[htbp]
  \centering
  \caption{Descriptive statistics of the ORIE network compared to that of Computer Science departments and Schools of Business. Assortativity is a measure of how much vertices with similar values of an attribute connect together (the Minimum Violation Ranking (MVR) is explained in section 4.)}\vspace{-0.5cm}
    \begin{tabular}{lcccc|cc|c}
          &       &        &      &       &\multicolumn{2}{|c|}{Assortativity}&  \\
       Network   & Vertices & Edges (\% Female) & Self-edges/edges & Density &     Degree& MVR & Reciprocity \\
       \hline
    ORIE  & 83    & 1179 (19.5)  & 0.1399 & 0.1121 &  0.1452 & 0.4614  & 0.1538 \\
    CS & 205   & 4388 (25.6)  & 0.0711 & 0.0688 & 0.2964 & 0.5327 &  0.1264 \\
    Business & 112   & 7856 (16.8)  & 0.0556 & 0.2738 & 0.2661 &  0.4330 & 0.2197 \\
    \end{tabular}%
  \label{tab:1}%
\end{table}
\end{small}
Two distinctive characteristics of the ORIE network relative to CS and Business (see Table \ref{tab:1}) are {\em a)} its much larger proportion of ``incestuous" Ph.D. hiring, given by the self-edges in the network (14\% of all faculty, almost three times that in Business and close to twice that of CS) and {\em b)} its lower degree assortativity.  Figure \ref{fig:0} (top) depicts the number of self-hires among the 83 ORIE departments, sorted in a hierarchy that will be explained below. Note how prevalent self-hiring is: 74\% of the ORIE departments have (as of summer 2016) at least one former Ph.D. student hired as faculty. This is similar than in Business schools (74\%) but higher than in Computer Science departments (58\%). Almost 14\% of all ORIE faculty are hired by their originating department. In some departments, self-hires account for more than 40\% of the faculty (see Figure \ref{fig:0}, bottom). 
 Our dataset, however, lacks information about cases when a faculty returns to her  alma mater after working in other departments, that is, cases where the self-hiring was not direct from graduate school to faculty. 

The {\em assortativity} index is a correlation measure of how much vertices with similar values of a given scalar attribute tend to be connected together (for this and other descriptive network statistic definitions, see, e.g., \citet{Newman}). Shown in Table \ref{tab:1} are  the assortativity measurements with respect to two vertex attributes. The first one, assortativity based on the total degree of each vertex (department) indicates a lower tendency for ORIE departments, compared to those in CS and Business, to establish hiring-placement connections with departments with a similar total number of hires and placements. The assortativity with respect to the MVR indices is an indication of how likely departments establish hiring-placement relations with departments equally ranked in a hiring hierarchy to be discussed in section \ref{Sec:3}. For ORIE, this assortativity is between that of CS and Business. 

Other descriptive statistics in the ORIE network shown in Table \ref{tab:1} tend to be also somewhere between those of the CS and the Business networks. The density or connectance of the ORIE network, defined as the number of non-zero entries in the adjacency matrix $\bf Y$ divided by the maximum number of possible edges, is between CS (which has a very sparse network) and Business (which is densest). As it will be shown below, the ORIE faculty hiring network is actually quite sparse except for a group of departments, those at the top of a hiring hierarchy, which connect more often among them. The proportion of female faculty in ORIE is 19.5\%, higher than in CS (16.8\%) but lower than in Business schools (25.6\%), an inequality that calls for further analysis that is beyond the present study.  Also shown in Table \ref{tab:1} is the {\em reciprocity}, the proportion of departments with exchanges of mutual hires.
The reciprocity of the ORIE network is also between that of CS and Business.  

\begin{figure}
\begin{center}
\resizebox{13cm}{9cm}{\includegraphics{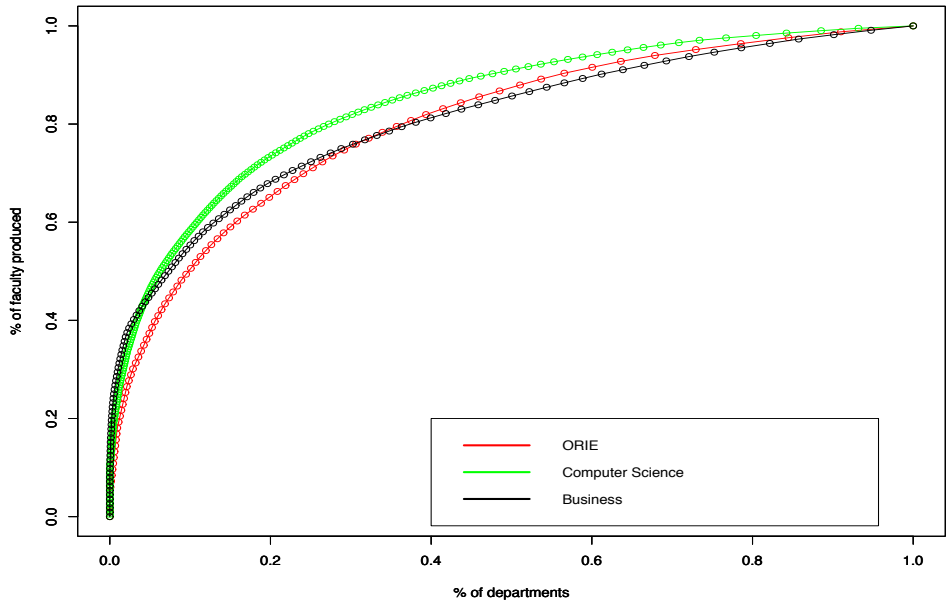}}
\caption{Lorenz curves indicating the inequality in the percentage of faculty production in ORIE, Computer Science, and Business as a percentage of the number of departments. About 10\% of ORIE departments generate close to half of all ORIE faculty.}
\label{fig:1} 
\end{center}
\end{figure}

Finally, the inequality in the production of Ph.D.'s among ORIE departments, although quite large, is also between that of CS departments and Business schools. Figure \ref{fig:1} shows the Lorenz curves for the proportion of Ph.D.'s produced. Approximately, 10\% of the ORIE departments generate about 50\% of the faculty, although this is not as pronounced as in Business schools, which have a very steep Lorenz curve near zero, with around 3\% schools generating 40\% of the faculty.

\section{Existence of a linear dominance hierarchy}
\label{three}

Given a faculty hiring network, a question of interest is if it is possible to order the departments in such a way to form a {\em linear dominance hierarchy}. The concept of dominance, popular in social networks in ecology and competitive sports, refers to an attribute of the pattern of repeated interactions between two individuals, characterized by a consistent outcome in favor of the same individual in a dyad \citep{deVries1995}. The consistent winner is dominant and the loser subordinate. Individuals form a {\em linear} dominance hierarchy if and only if a) for every dyad $(i,j)$  either $i$ dominates $j$ or $j$ dominates $i$ and b) every triad is transitive, i.e., for any individuals $i$, $j$ and $k$ it is true that if $i$ dominates $j$ and $j$ dominates $k$, then $i$ dominates $k$. Transitivity is equivalent to a dominance hierarchy that is acyclic \citep{Ali1986}. 
 The notion of a dominance hierarchy has been studied in faculty networks by \citet{Clauset}, \citet{Huang} and \citet{LiuGonzalez}. Applied to faculty hiring networks, if a department $v_j$ hires more Ph.Ds from department $v_i$ than the number $v_i$ hires from $v_j$, the relationship between these two vertices implies a ``dominance" relation of department $i$ with respect to department $j$, given that  $j$ wishes more strongly to have access to the students produced in department $i$ and not viceversa (ties are therefore allowed). In his study of animal societies, \citet{Landau} defined a score structure $W = (w_i,w_2,...,w_n)$ where each $w_j$ equals the number of individuals dominated by element $j$. A {\em hierarchy} is then a score structure
\[
W = (w_1=n-1, w_2=n-2, ..., w_n=0)
\]
so that the members of the society can be ordered as $1 \succ 2 \succ 3 \succ... \succ n$, with each dominating all the members below it, and being dominated by all those above. At the opposite extreme is an ``egalitarian" society where assuming $n$ odd, $W=(w_1=\frac{n-1}{n}, w_2=\frac{n-1}{n},...,w_n=\frac{n-1}{n})$.  \citet{Landau} then introduced his ``hierarchy index" $h$, a measure of the variability of the $w_j$'s normalized so that $h=0$ implies equality (egalitarian dominance) and $h=1$ implies a perfect linear hierarchy, with $h$ equal to
\[
h = \frac{12}{n(n^2-1)}\sum_{j=1}^n \left( w_j - \frac{n-1}{2}\right)^2.
\]
Of course, a perfect linear hierarchy may not exist in a society, and some violations to a perfect linear hierarchy may exist, a topic we discuss in section \ref{Sec:3}. 

A practical difficulty when determining if a linear hierarchy exists in a society (and in general, for any other inference one desires to attempt based on observed social interactions) is that we may have few data points about individuals interacting, with many not interacting during the observation period. The basic approach to deal with the spareness of the observed adjacency matrix (or sociomatrix) $\bf Y$  is to treat it is a realization of a stochastic process, and to use nonparametric tools for statistical inference. Along this approach is \citet{deVries1995}, who introduces a randomization test for the hypothesis of no linear hierarchy based on Landau's $h$ statistic that takes into account unknown tied relationships (in our case, these are departments that have never hired each other's Ph.D. students). To perform the test, each dyad ($i,j$) is randomized $m$ times and the $h$ statistic is computed for each random sociomatrix giving a set of numbers $\{h_s\}_{s=1}^m$. $h$ is also computed for the observed sociomatrix. If the empirical p-value of the test, defined as $1- |\{h_s > h\}|/m$ (where $| \cdot |$ indicates cardinality) is small, this is evidence the observed linear hierarchy is statistically different than that of a random matrix, which has no hierarchy. 

\begin{figure}
\begin{center}
\begin{tabular}{ccc}
\resizebox{5cm}{4cm}{\includegraphics{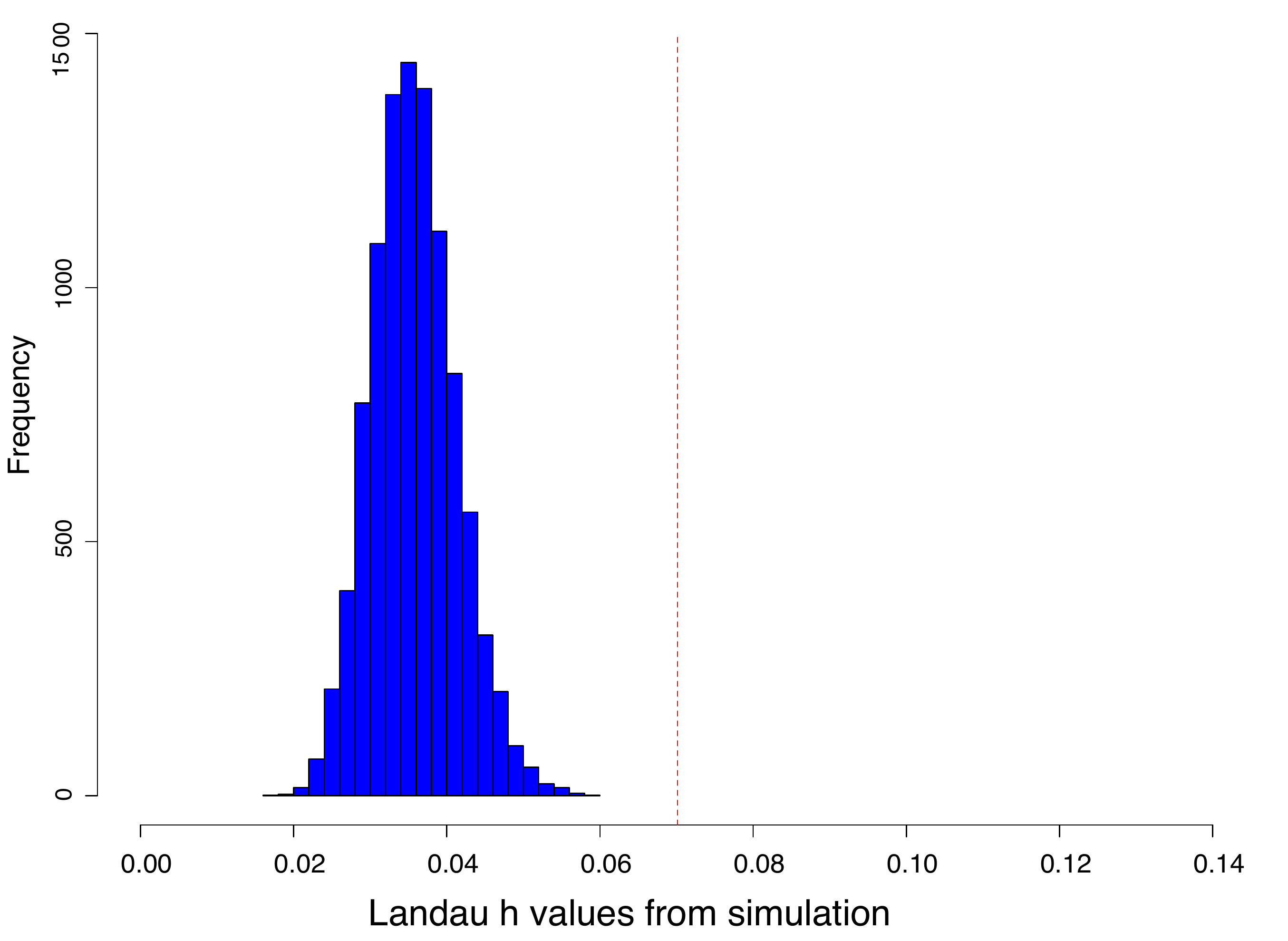}}&
\resizebox{5cm}{4cm}{\includegraphics{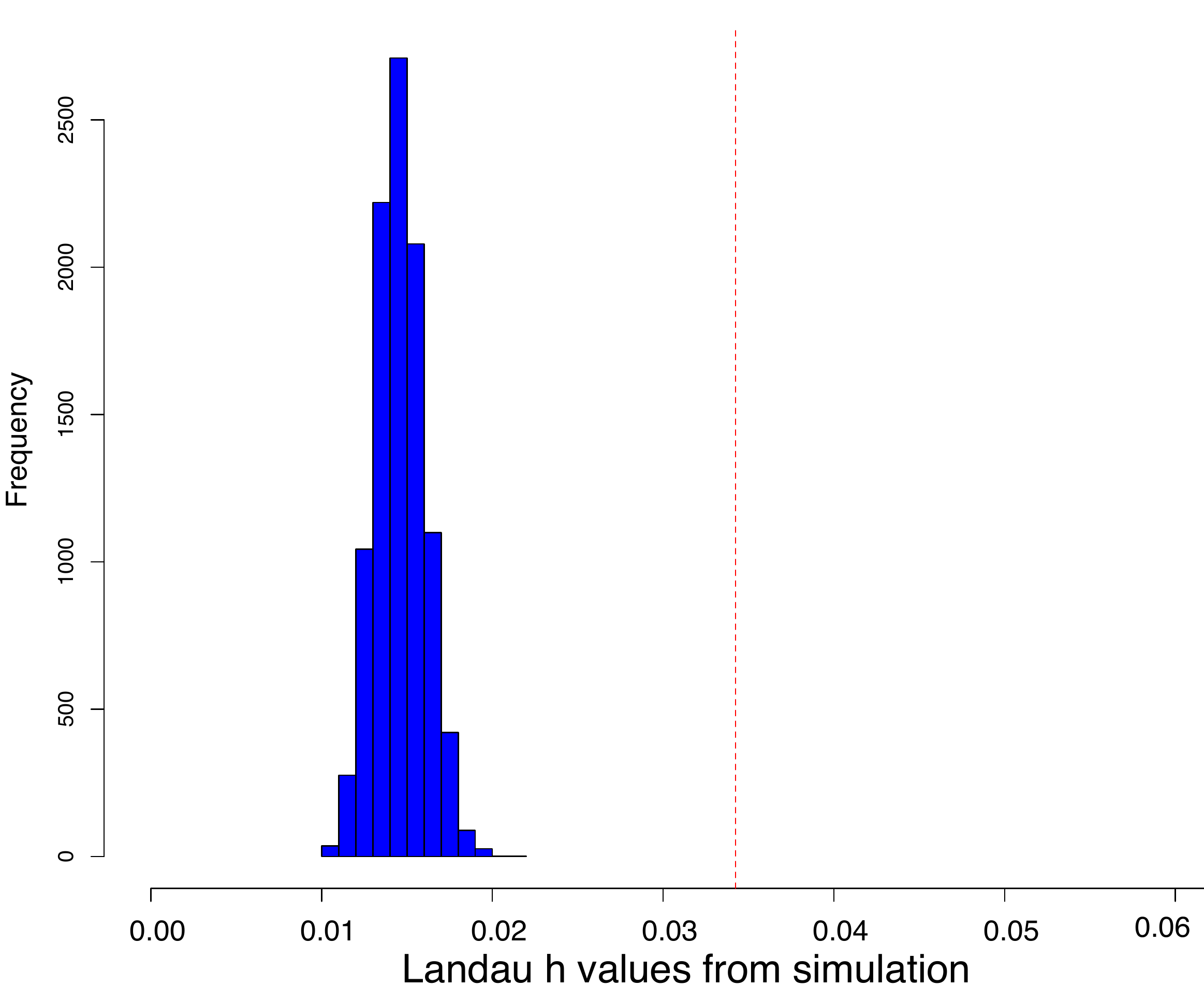}}&
\resizebox{5cm}{4cm}{\includegraphics{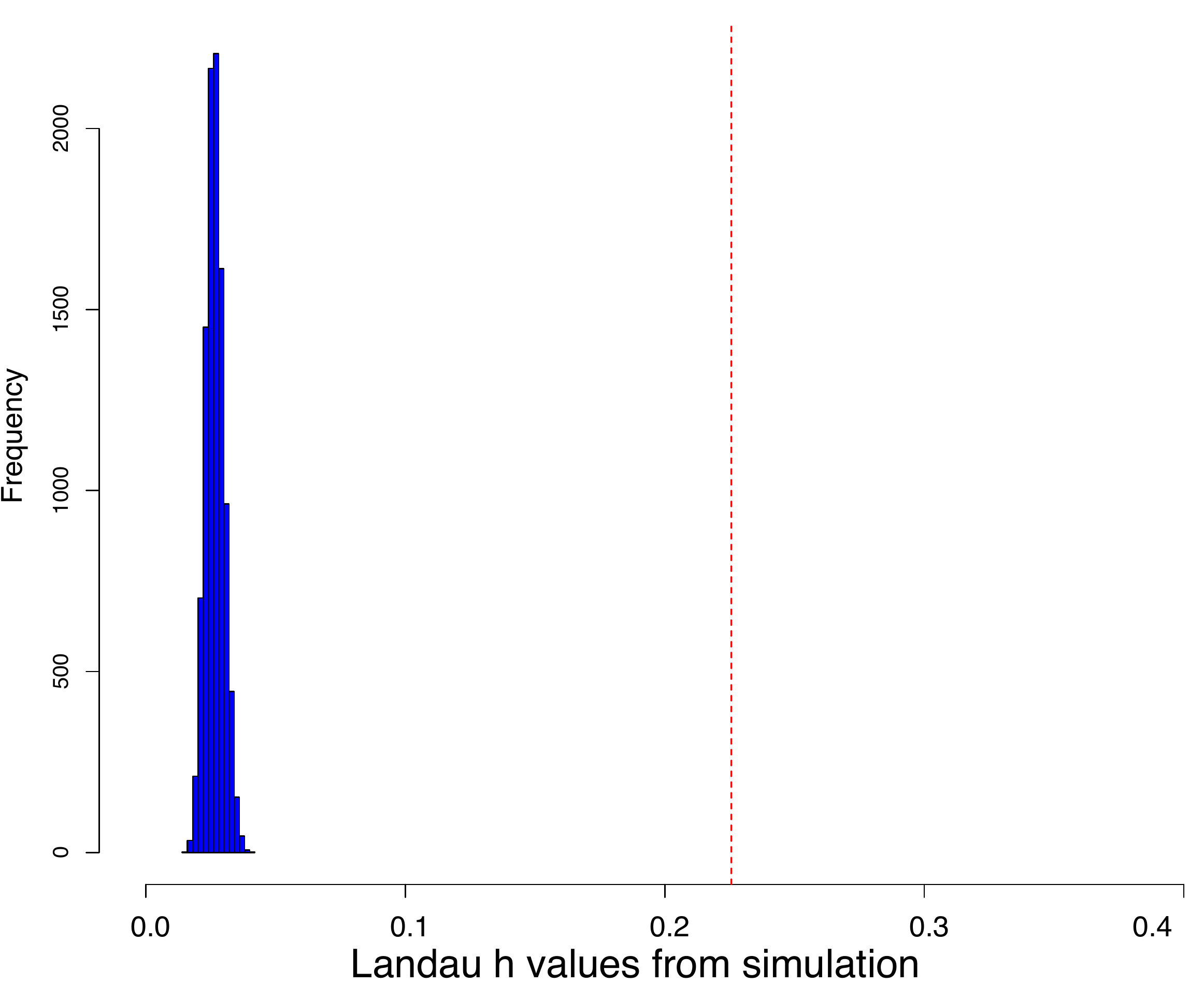}}\\
\end{tabular}
\caption{DeVries' test for the presence of a linear dominance hierarchy in the complete $(n=83)$ ORIE network (Left), the Computer Science network (Middle, $n=205$) and the Business network (Right, $n=112$). Randomization distribution (10K simulations) under the null hypothesis of random graph of Landau's $h$ values. Red vertical lines are the observed $h$ values. In all cases, the empirical p-values are 0.0, indicating strong evidence of the presence of a hierarchy in each of three faculty networks. }
\label{Fig:2}
\end{center}
\end{figure}

Ordering individuals in a linear or near-linear hierarchy is justified only if there is statistical evidence in favor of such hierarchy comparing the existing hierarchy to what could be obtained from a random matrix. Finding a near-linear hierarchy ($h<1$) is a hard combinatorial problem about which we comment below. 
Figure \ref{Fig:2} shows results of the deVries tests applied to the ORIE, CS and Business networks (data for CS and Business are from 2015 taken from \citet{Clauset}). The empirical p-values are zero for all 3 disciplines, implying there is a statistically significant linear hierarchy in each of the three networks.

If a linear or near-linear hierarchy is significant, a related question is how steep is the hierarchy. In Appendix 2 we statistically test for the significance of the steepness of a linear hierarchy \Citep{deVries2006}, and find that the ORIE, CS and Business networks all have a hierarchy with a statistically significant slope. 

The tests for the significance of a linear hierarchy and for a significant slope were repeated for the top ORIE departments (sorted according to the MVS$_2$ index in Table A1 in Appendix 1, further discussed below). Table \ref{tab:2} shows the results of these tests applied to an increasing number of (sorted) departments.  The slope is quite steep and significant only for the first 10 departments. Overall, for the 83 ORIE departments the slope is statistically different from zero, but its actual numerical value is rather small.

Similarly to the tests in this section, in the sequel we will again treat the observed network as a noisy realization of possible networks, and following \citet{Clauset} we will use a ``bootstrapping" approach to network creation to help perform an analysis that considers other possible (sparse) networks that could have been observed by chance.

\begin{table}[h]
  \centering
  \caption{Dominance hierarchy linearity and steepness test results for different ORIE subnetworks. The hierarchy is steeper closer to the top, where it is also much  denser. Hierarchy considered is the MVS$_2$ hierarchy explained in section \ref{Sec:3}.}
    \begin{tabular}{cccccc}
    No. of departments    & Density(connectance)&Observed $h$ & p-value & Slope & p-value \\
    \hline
    Top 10  & 0.6000&0.5590 & 0.0307 & -0.4281 & 0.0000 \\
    Top 20  & 0.4025&0.3036 & 0.0098 & -0.2061 & 0.0000 \\
    Top 50  & 0.1912&0.1328 & 0.0004 & -0.0688 & 0.0000 \\
    All 83  & 0.1121&0.0701 & 0.0000 & -0.0278 & 0.0000 \\
    \end{tabular}%
  \label{tab:2}%
\end{table}
\subsection{A caveat: ``observational zeroes" in a sparse sociomatrix}
\label{zeroes}
The preceding analysis indicates that the 83 ORIE departments can be ranked according to a statistically significant near linear hierarchy using hire-placement data, even though the hierarchy appears to be quite flat for most of the $n$ departments except at the very top of the hierarchy. The de Vries' steepness test  (Appendix 2) depends on the number of interactions between the actors or individuals in the network. In animal behavior, the ideal way to find a hierarchy is to observe pairwise competitions in a balanced (designed) ``tournament" \citep{David}, but in the field of social network data an abundance of ``observational zeroes" makes it more difficult to determine a hierarchy. Although no formal power analysis is available, \citet{deVries1995} gives some numerical evidence to suggest that the power of the linearity test to detect an existing hierarchy goes down when the frequency of observational zeroes increases, and it is quite possible the same occurs in the steepeness test. Table \ref{tab:2} shows how the connectance (density) of the departments among the top of the hierarchy is much higher than among the rest, and this implies the test statistics have more information about placements and hires within this group that among dyads that include at least one individual ranked lower in the hierarchy. 

The predominance of zeroes in a sociomatrix may not necessarily indicate there are no dominance relations among the dyads. As warned by \citet{deVries2006} one should be careful when interpreting linearity and steepness tests for societies in which there are very few pairwise interactions recorded. This warning is by extension worth keeping in mind when trying to find a dominance relation or ranking in a faculty hiring network. We can distinguish between three types of ``observational zeroes" in the adjacency matrix (or sociomatrix): a) in animal behavior, if two individuals are not observed to interact, it may be because there is a dominance relation present, with the subordinate individual avoiding the dominant individual. We will call the similar case of departments avoiding hiring from higher ranked departments {\em deference} or {\em avoidance} zeroes. Alternatively, b) there may simply be no dominance-subordinate relation between two individuals that ``respect" each other,  a case analogous to departments that do not have hiring-placement relations simply due to 
their finite capacity, which limits the possibility of observing more hiring-placement relations, a case we will call {\em equality} zeroes. Finally, c) it could happen that we have not observed long enough the network and the question of whether there is a dominance between two individuals is unresolved. This would be the case when relatively young departments have not established hiring-placement relations with many departments simply because their young age. This case will be called {\em unresolved} zeroes. It is not possible, however, to determine from the data what kind of ``observational zeroes" one is dealing with in a particular hiring network. 

With these precautionary notes we now proceed to find a near linear hierarchy in the ORIE network. For the CS and Business networks, see \citet{Clauset}. The presence of ``observational zeroes" and the low density of the ORIE network justifies the bootstrapping approach of the next section. Likewise, the different connectance among strata of the hierarchy and its corresponding changes in steepness also justifies the search for groups (communities) of departments, a task we address in Section \ref{Sec:5}.

\section{The Minimum violations and minimum violations and strength indices}
\label{Sec:3}
Let ${\bf Y}$ be the adjacency matrix of a faculty hiring network. \citet{Clauset} define a hierarchical index given by a permutation $\pi$ that induces the minimum number of edges that ``point up" the hierarchy. This is found by
\be
\min_{\pi({\bf Y})} \; \; S(\pi({\bf Y})) = \min_{\pi({\bf Y})} \; \mathop{\sum \sum}_{i > j} Y_{ij}\; \mbox{sign}(\pi(i) - \pi(j))
\label{MVR}
\ee
 Thus, if there is an edge from $i$ to $j$ $(Y_{ij} >0)$ and $\mbox{sign}(\pi(i) - \pi(j)) = +1$ (i.e., $\pi(i) > \pi(j)$) this means a lower ranked department in the hierarchy  has placed a faculty at a higher ranked department (recall rank one is highest), in other words, we have a ``violation" of the hierarchy implied by $\pi$ and this will increase $S(\pi({\bf Y}))$. Similarly, if there is an edge from $i$ to $j$ and $\mbox{sign}(\pi(i) - \pi(j)) = -1$ (i.e., $\pi(i) < \pi(j)$), this is not a violation, and will make $S(\pi({\bf Y}))$ decrease. A permutation $\pi({\bf Y})$ that minimizes (\ref{MVR}) is called a {\em minimum violation ranking} (MVR) which has been proposed as the optimal way of rankings players in a round-robin tournament \citep{Ali1986}. For networks where a perfect linear hierarchy does not exist ($h<1$, thus violations are unavoidable) solving (\ref{MVR}) is a hard combinatorial problem.  Problem (\ref{MVR}) is in particular equivalent to reordering the columns and rows of the adjacency matrix such that we get an upper triangular matrix, a problem which has been proved to be NP-complete \citep{Charon2010}. For the 83 departments in the ORIE network, however, an exact algorithm for finding the MVR's, such as  \cite{Pedings} binary linear integer programming formulation, is computationally feasible. Note that multiple optimal MVR rankings with the same number of violations may exists in a complex network. Although we will argue that the MVRs do not fully reflect the hierarchy of a faculty hiring network,  Table \ref{tab:A1} in Appendix 1 lists the MVR rankings of the 83 ORIE departments considered in this study for completeness. 

Rather than solving (\ref{MVR}), which considers only the number of violations, we could also consider the {\em strength} of each violation.  To obtain a dominance relation in a society of animals, \citet{deVries1998} defines the strength of the violations in a sociomatrix as:
\be
\min_{\pi({\bf Y})} \; \; S_{\mbox{\tiny MVS}_1}(\pi({\bf Y})) = \min_{\pi({\bf Y})}\mathop{\sum \sum}_{i>j} (i-j)Y_{ij}
\label{MVS1}
\ee
where the difference $(i-j)$ (with $i>j$) measures the strength of a violation in the ranking (which exists if $Y_{ij}>0$). In our case, entries under the diagonal are {\em unexpected faculty hires}, where department $j$ hires a number of faculty from a lower ranked department $i$. We will refer to the resulting rankings from  solving (\ref{MVS1}) the {\em Minimum violations and strength} rankings and will denote them by MVS$_1$. To obtain these rankings, we modify the stochastic search algorithm in \citet{Clauset} to account for the strength of a violation (see Appendix 3). The ORIE MVS$_1$ rankings are shown in Table \ref{tab:A1} in Appendix 1.\vspace{0.1cm}

\begin{figure}[htbp]
\begin{center}
\begin{tabular}{lr}
\resizebox{8cm}{8cm}{\includegraphics{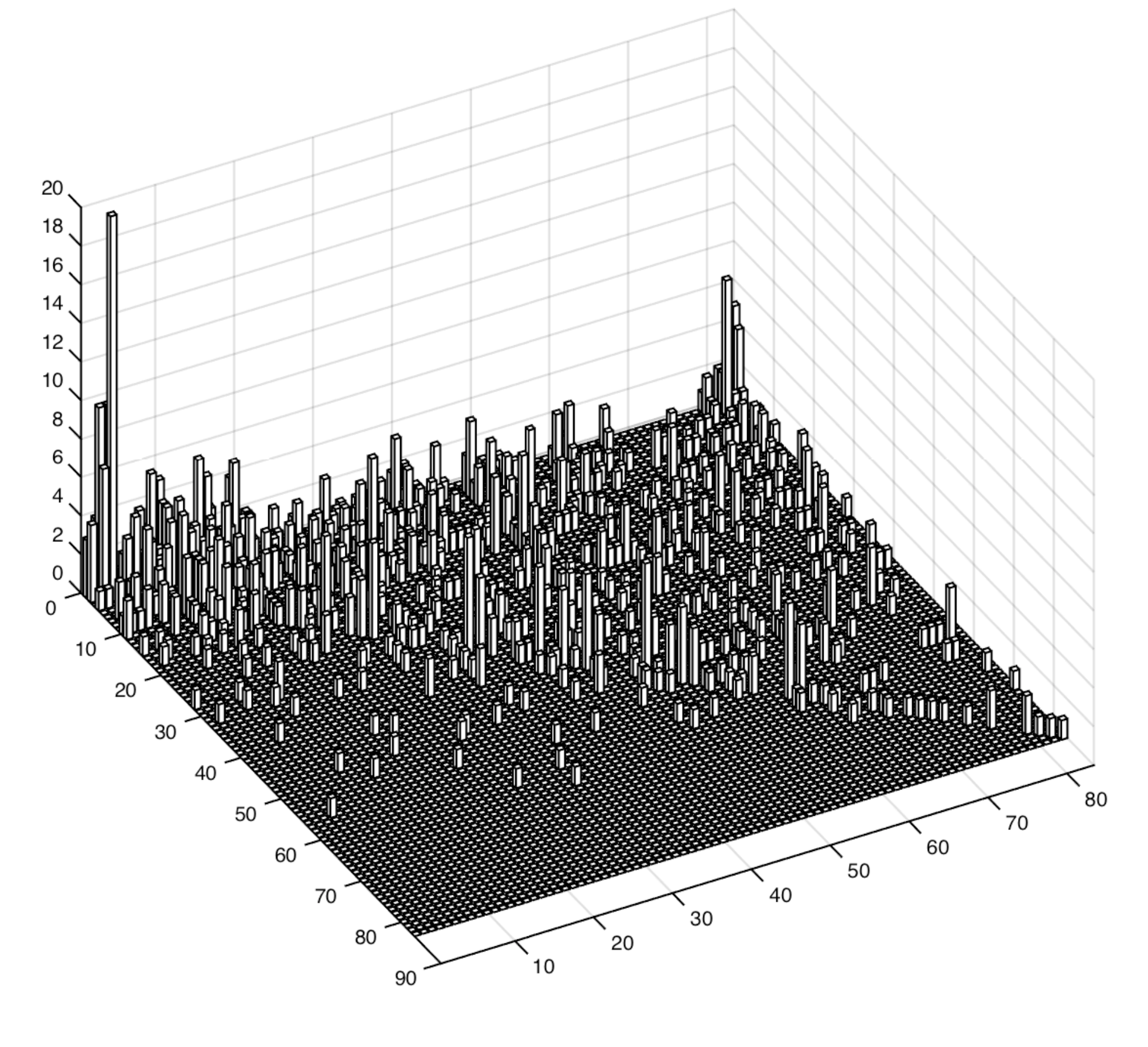}}&
\resizebox{8cm}{8cm}{\includegraphics{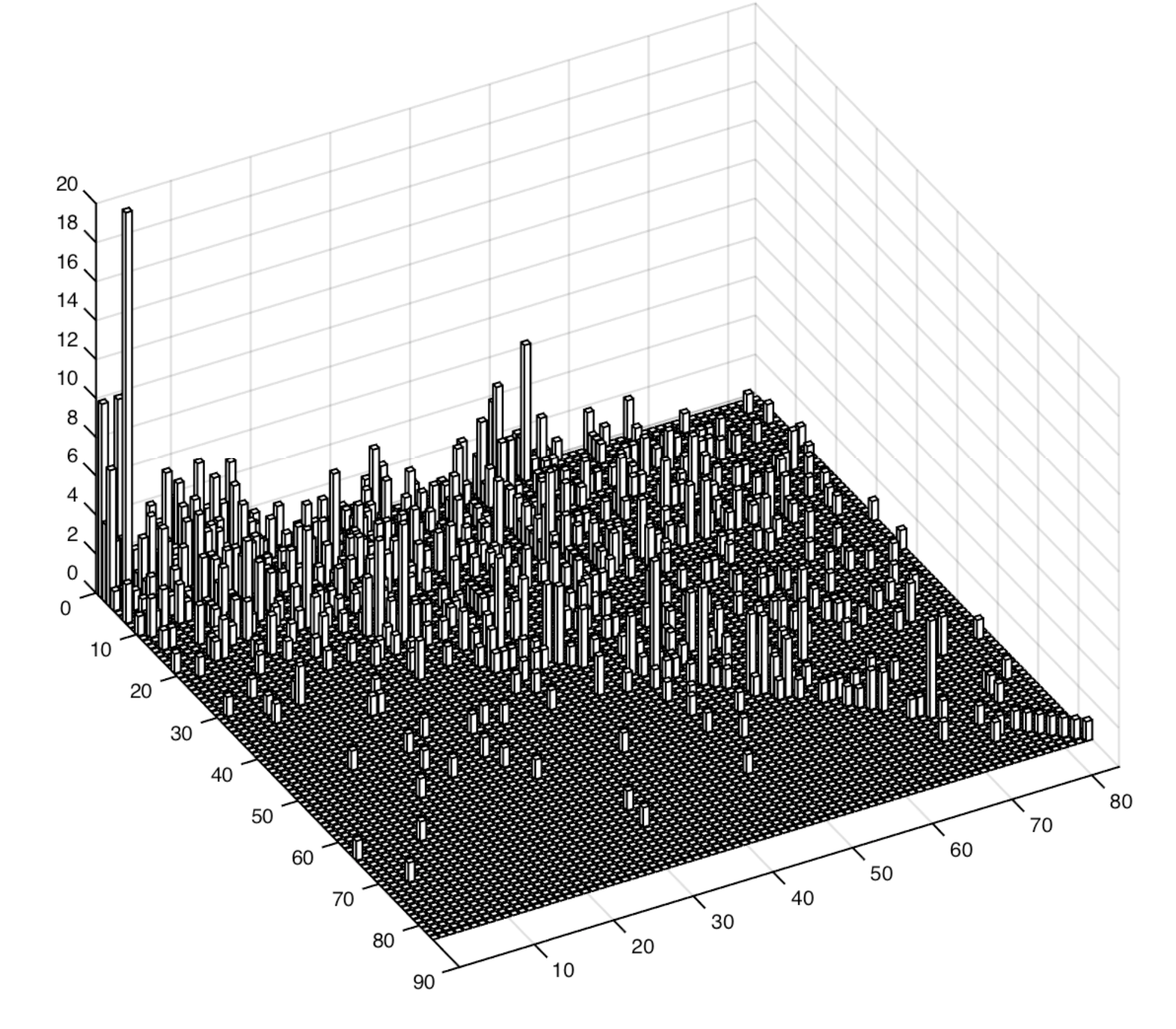}}
\end{tabular}
\caption{Left: matrix bar plot of the ${\bf Y}$ adjacency matrix under the MVS$_{1}$ rankings, which considers the strength of a type 1 violation (unexpected placement of a faculty from a lower to a higher ranked department). Right: matrix bar plot of the ${\bf Y}$ adjacency matrix under the MVS$_{2}$ rankings, which considers the strength of a type 1 violation and also of type 2 violations (unexpected number of hires from lower ranked departments). The plot on the right is closer to ``hillside" than the one on the left, a consequence of the different objectives. Note the peak in the upper right corner in the plot on the left, these are departments that hire an unexpected number of faculty from the top departments, yet ended up, under the MVS$_{1}$ rankings, in a very low position. The MVS$_{2}$ rankings ameliorate this situation to a certain degree.}
\label{fig:hillside}
\end{center}
\end{figure}

A problem with both the MVR and MVS$_1$ rankings previously proposed in the literature on hiring networks is that a department that places few of its own Ph.D. students, despite hiring from the top departments, may be ranked very low. For instance, this is the case, under both MVR and MVS$_1$ rankings, of Naval Postgraduate School OR dept., which received among the lowest rankings using these two indices. 
A department that consistently hires from top ranked departments should be  ranked higher than one that not only does not place its Ph.Ds but also does not have any hiring interactions with the top ranked group. To illustrate, a bar graph of the adjacency matrix $\bf Y$ sorted according to the MVS$_1$ rankings indicates the nature of the problem (Figure \ref{fig:hillside}, left): while both (\ref{MVR}) and (\ref{MVS1}) tend to minimize the number of entries below the diagonal, the entries above the diagonal are not considered. Note the large $Y_{ij}$ values in the upper right corner of the matrix plot; these correspond to low ranked departments under the MVS$_1$ criterion that hired repeatedly from the very top departments in the ranking, yet they ended up with a very low MVS$_1$ ranking. To correct this anomaly, we propose to also account for this type of secondary violation, which we will call {\em unexpected placements} i.e., when a top department places too many Ph.Ds in other lower ranked departments that should not be that attractive for their graduating students. Penalizing this kind of violation (and its strength) will result in an improved ranking for a department that ``hires high". A compound criteria, including the strength of both unexpected hires and unexpected placements is:
\be
\min_{\pi({\bf Y})} \; \; S_{\mbox{\tiny MVS}_2}(\pi({\bf Y})) = \min_{\pi({\bf Y})}\mathop{\sum \sum}_{i>j} (i-j)Y_{ij} + (i-j) Y_{ji}
\label{MVS2}
\ee
where $Y_{ji}$ in the second term considers abnormal entries above the diagonal (since $i>j$) and $(i-j)$ measures the strength of this type of violation, similarly to (\ref{MVS1}).  It would be wrong to use this compound criteria as it gives equal weight to the two types of violations, whereas unexpected hires (or violations, i.e., entries below the diagonal) should be accounted more severely than unexpected placements (above the diagonal). Assigning weights to each objective is {\em ad-hoc} and therefore not a solution. Instead, to give primary importance to violations (unexpected hires), the pairwise stochastic swapping algorithm that we utilize (Appendix 3) in solving (\ref{MVS1}) was defined to accept a new ranking (i.e., exchanging the rankings of two departments in question) if either:
\begin{enumerate}
\item the number of violations is lower after the switch, or
\item the number of violations is the same as before the switch, but the sum of the strengths of the two types of violations, $S_{\mbox{\tiny MVS}_2}(\pi({\bf Y}))$ is lower after the switch.
\end{enumerate}

We refer to the resulting rankings as MVS$_2$ rankings. They are reported in Appendix 1, Table \ref{tab:A1}.

A concept related to the two types of violations, unexpected hires and unexpected placements, found in the area of sport teams rankings, is that of an adjacent matrix in ``hillside" form (\cite{Pedings}). In sport team rankings, the adjacency matrix contains the point or goal differential between all pairs of teams in a tournament. A sports tournament sociomatrix ${\bf Y}$ is in hillside form if it is ascending along all rows and descending along all columns, i.e., if
$ Y_{ij} \leq Y_{ik},  \; \forall i, \forall j \leq k, $ and $Y_{ij} \geq Y_{kj},  \; \forall j, \forall i \leq k$ \citep{Pedings}.
Associated with this definition there are two types of violations: ``upsets", nonzero entries below the diagonal matrix, and ``weak wins", entries above the diagonal matrix that do not follow a hillside pattern, i.e., when team $i$ did not score as many points as it would have been expected when playing team $j$, with $j>i$.

While the concept of weak wins and upsets in a sports tournament is similar to that of unexpected hires and placements, there are important differences. In a sports tournament, the sociomatrix contains goal differentials, and therefore the hillside form as defined above is the ideal form of a hierarchy. Contrary to a tournament, where every team plays against all other teams and therefore a {\em complete comparison network} can be formed, we do not have pairwise comparisons for all dyads in a faculty hiring network, i.e., not all edges are observed as we have ``observational zeros" as mentioned before, and this requires special consideration.  In addition, in a faculty hiring network the definition of a {\em hillside form} adjacency matrix needs to be modified to account for both unexpected hiring and unexpected placements since we wish to find a hierarchy such that we have both descending columns {\em and descending rows} as well:
\[\vspace{-0.17cm}
Y_{ij} \geq Y_{ik},   \quad \forall i, \forall j \leq k, \quad \mbox{and} \quad Y_{ij} \geq Y_{kj},  \quad \forall j, \forall i \leq k.
\]
That is, instead of winning by more goals against decreasingly ranked opponents, higher ranked departments are expected to place fewer faculty at decreasingly lower ranked departments.
 The MVS$_1$-ranked matrix (see Figure \ref{fig:hillside}, left) is not in what we define as hillside form since there is a ``peak" on the upper right cell of the matrix.\vspace{0.1cm}
 
 Figure \ref{fig:hillside} (right) shows the adjacency matrix for the ORIE departments sorted according to the MVS$_2$ indices. Note how the matrix is closer to hillside form, i.e., a matrix with both rows and columns closer to being monotonically descending (Naval Postgraduate School then deservedly ranks much higher, and the peak on the left plot is moved considerably further to the left of the matrix).

\subsection*{Bootstrapping the observed network}

\begin{figure}[H]
\begin{center}
\resizebox{15cm}{9cm}{\includegraphics{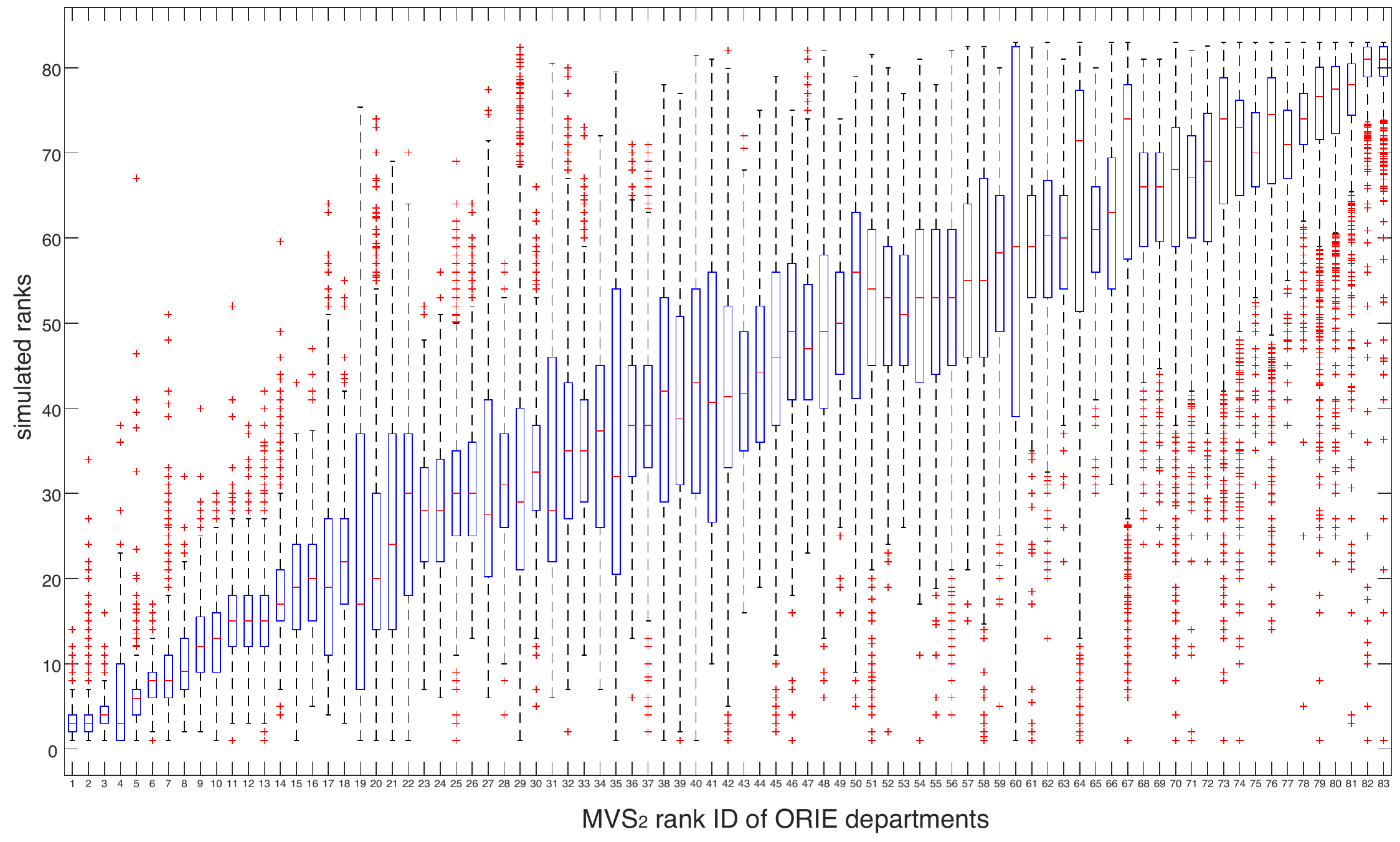}}
\caption{Boxplots of the boostrapped Minimum Violation and Strength (MVS$_{2}$) ranks for all 83 departments in the ORIE network, sorted by mean MVS$_2$ value (1000 bootstrapped networks). The reported MVS$_2$ ranks correspond to the average values. Note how the extreme ranks are less uncertain, a characteristic also reported for other disciplines \citep{Clauset}.} 
\label{fig:bootNet}
\end{center}
\end{figure}

\begin{figure}
\begin{center}
\resizebox{12cm}{10cm}{\includegraphics{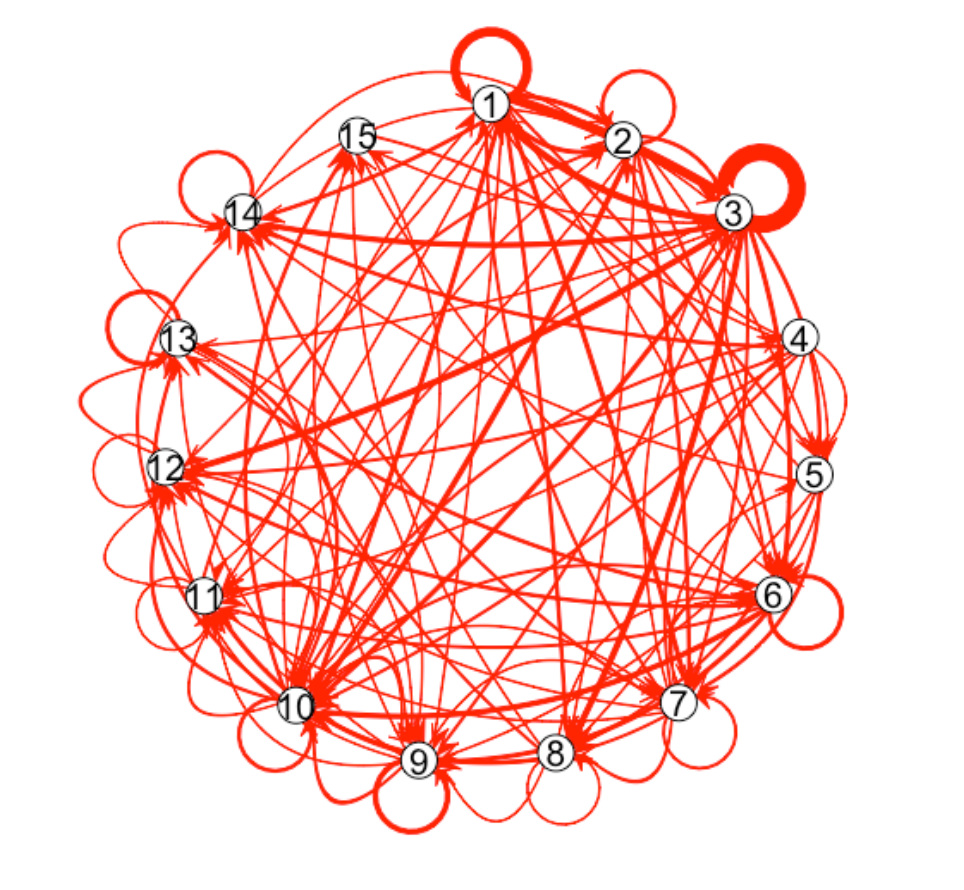}}
\caption{ORIE faculty hiring network for the top 15 departments sorted according to MVS$_2$ index. Edge width is proportional to number of hires. The edge density (or connectance) of the ORIE network is high only for the departments at the top of the hierarchy, but otherwise it is quite sparse, see Table \ref{tab:2}. } 
\label{fig:top15}
\end{center}
\end{figure}

Given the sparseness of the ORIE network due to observational zeroes, we provide more robust MVS$_2$ (and MVS$_1$) indices by optimizing 1000 ``bootstrapped", or randomly sampled (with replacement) ORIE networks. The bootstrapped networks are obtained by sampling edges, with replacement, from the observed ORIE network, such that the probability of sampling an edge is proportional to the number of faculty $Y_{ij}$ in that edge. Each bootstrapped network ${\bf Y}_b$ was then optimized according to the $S_{\mbox{\tiny MVS$_1$}}(\pi({\bf Y}_b))$ or  $S_{\mbox{\tiny MVS$_2$}}(\pi({\bf Y}_b))$ criteria using the stochastic search method of Appendix 3, and the resulting MVS values were recorded for each department. The results for MVS$_2$ across the ensemble of bootstrapped replications are shown in Figure \ref{fig:bootNet}. Note how there is less uncertainty in the MVS$_2$ values in the extremes of the hierarchy, and more uncertainty for those ranked in the middle, a phenomenon also reported for CS and Business networks \citep{Clauset}.  The final indices MVS$_1$ and MVS$_2$ reported on Table \ref{tab:A1} are the average ranks computed from the 1000 bootstrapped and optimized networks. A network representation of the top 15 MVS$_2$ departments is shown in Figure \ref{fig:top15}. This is the most dense part of the ORIE network.

\subsection*{Correlations between different ranks}

Table \ref{tab:corrs} shows the Spearman correlation coefficients between various rankings computed for the ORIE departments, including two published rankings (NRC and US News \& World Report, see Appendix 1) and the MVR and MVS$_2$ rankings described earlier. All these rankings are shown in Table \ref{tab:A1} in the Appendix 1. Not surprisingly, the MVR and MVS$_2$ rankings are highly correlated, but a more unexpected high correlation is between the MVS$_2$ rankings and ``Hub" importance ranks. In a social network, {\em Hubs} and {\em Authorities} are types of vertices defined in an intertwined manner: a Hub is a vertex that points to many other vertices with high authority, and an authority is a vertex pointed to by many hubs \citep{Newman}. In a faculty hiring network, vertices with high hub ranking are departments that ``feed" faculty to departments sought after by faculty candidates, thus Hub importance refers to {\em placement capacity} to departments that in turn are important (similarly, the authority ranks refer to {\em hiring capacity}, the ability to attract and hire faculty from important departments). 
MVS$_2$ is also correlated with the out-degree rankings, but the correlation is not as high as with hub importance: a department must not only generate many Ph.Ds but must place them in the best possible departments. The PageRank, (left) Eigenvalue, and out degree rankings are all highly correlated because the importance of ``centrality" in a faculty hiring network is bestowed by out degree.

While there is significant correlation between published rankings and the MVS$_2$ and MVR indices, there are significant differences especially among the top 20 departments (see Appendix 1 Table \ref{tab:A1}). For the top 20 MVS$_2$ departments, the Spearman correlation is 0.796 and 0.711 between the MVS$_2$ and US News and NRC rankings, respectively, with two departments in the top 20 MVS$_2$ indices not in the US News list.


\begin{scriptsize}
\begin{table}
 \caption{Spearman correlation coefficients between the ranks of common measures of vertex importance, including public rankings, in the observed ORIE network. Only pairs of entries with no NA values were considered. (USN = US News 2016 ranks, NRC = 2011 NRC ranks).}
    \label{tab:corrs}

    \begin{tabular}{l|rrrrrrrrrrr}
    \textbf{} & \textbf{MVS$_2$} & \textbf{MVR} & \textbf{USN} & \textbf{NRC} & \textbf{In-deg} & \textbf{Out-deg} & \textbf{Eigen} & \textbf{PgRank} & \textbf{Bet.} & \textbf{Hub} & \textbf{Auth.} \\
    \hline
    \textbf{MVS$_2$} & 1.000 &       &       &       &       &       &       &       &       &       &  \\
    \textbf{MVR} & 0.921 & 1.000 &       &       &       &       &       &       &       &       &  \\
    \textbf{USN} & 0.857 & 0.776 & 1.000 &       &       &       &       &       &       &       &  \\
    \textbf{NRC} & 0.849 & 0.819 & 0.876 & 1.000 &       &       &       &       &       &       &  \\
    \textbf{In-deg} & 0.521 & 0.275 & 0.690 & 0.499 & 1.000 &       &       &       &       &       &  \\
    \textbf{Out-deg} & 0.884 & 0.832 & 0.825 & 0.773 & 0.585 & 1.000 &       &       &       &       &  \\
    \textbf{Eigen} & 0.873 & 0.840 & 0.757 & 0.789 & 0.438 & 0.863 & 1.000 &       &       &       &  \\
    \textbf{PgRank} & 0.861 & 0.860 & 0.799 & 0.775 & 0.461 & 0.942 & 0.938 & 1.000 &       &       &  \\
    \textbf{Bet.} & 0.490 & 0.418 & 0.479 & 0.518 & 0.539 & 0.754 & 0.659 & 0.709 & 1.000 &       &  \\
    \textbf{Hub} & 0.923 & 0.855 & 0.857 & 0.845 & 0.569 & 0.933 & 0.917 & 0.906 & 0.637 & 1.000 &  \\
    \textbf{Auth.} & 0.724 & 0.557 & 0.801 & 0.693 & 0.776 & 0.613 & 0.607 & 0.574 & 0.341 & 0.712 & 1.000\vspace{0.2cm}
    \end{tabular}
       \end{table}
    \end{scriptsize}

\section{Latent location variables and clustering of ORIE departments: a latent exponential random graph model}
\label{Sec:5}
Descriptive statistics as those in section \ref{Sec:2} can only provide partial information about the structure of a complex network. Single indices such as the MVR and MVS indices (or such as published rankings) that try to capture the ``prestige" of an academic department are inherently incomplete. One feature that was evident from the descriptive analysis of the ORIE network (Sections \ref{Sec:2} and \ref{Sec:3}) was that there is a core of departments at the top of the hierarchy that form denser connections, while the periphery is much more sparsely connected. Also, there is evidence the steepness of the hierarchy varies within the hierarchy (Appendix 2 and Table \ref{tab:2}). This indicates that in order to better understand the structure of the network, rather than finding a linear hierarchy based on single indices, it is worth finding groups of similar departments. 

In this section, rather then simply applying clustering algorithms directly to the ORIE network data, we first consider the observed ORIE network as a noisy realization, or sample, from a stochastic network model. We study a particular type of exponential random graph model (ERGM) in which the conditional probability of a tie between two actors, given covariates (if any), depends on the distance between actors in an unobserved or latent ``social space" \citep{latentnet}.

In a latent ERGM model, the sociomatrix (or adjacency matrix) $\bf Y$ is viewed as a random variable that depends on parameters $\bm \beta$, covariates $\bf x$ and positions ${\bf Z} = \{{\bf Z}_i \}_{i=1}^n$ in a `social space". The latent ERGM model we used is:
\begin{eqnarray*}
P(\bf Y = \bf y | {\bm \beta}, {\bf x, Z}) &=& \prod_{i,j} P(Y_{i,j}=y_{ij}|{\bm \beta}, {\bf x, Z}) \\
P(Y_{ij} = y_{ij} | {\bm \beta}, {\bf x},{\bf Z}) &=& f(y_{ij}|\mu_{ij}) \; \; = \; \; \frac{\exp(-\mu_{ij})\mu_{ij}^{y_{ij}}}{y_{ij}!}\\
\log(\mu_{ij}) = E[Y_{ij}|{\bm \beta}, {\bf x}, {\bf Z}] &=& \eta_{ij}({\bm \beta}, {\bf x}, {\bf Z})\\
\eta_{ij}({\bm \beta}, {\bf x}, {\bf Z}) &=& \sum_{k=1}^p x_{k,i,j} \beta_{k} - |{\bf Z}_i-{\bf Z}_j| 
\end{eqnarray*}
We thus adopted a Poisson density for the number of hires between two departments $Y_{ij}$ and a link function $\eta = g(\mu)=\log(\mu_{ij})$ which is log-linear in the covariates and in the distances in an Euclidean latent space (defined by the ${\bf Z}_i$'s). The latent positions are assumed to follow a mixture of spherical multivariate normals in the social space, which could have in principle any dimension $d$:
\be
{\bf Z}_i \; \iid \; \sum_{g=1}^G \lambda_g \mbox{N}_d({\bm \mu}_g, \sigma^2_g {\bf I}_d), \quad i=1,...,n
\label{Zs}
\ee
where $0 \leq \lambda_g \leq 1$ is the probability  an individual belongs to group $g$ ($\sum_g^G \lambda_g=1)$

Fitting the model implies finding estimates for the parameters ${\bm \beta}, (\lambda_g, {\bm \mu}_g, \sigma_g^2)$ for $g=1,...G$. A Bayesian formulation has been proposed by Krivitsky et al. \citep{Krivitsky} who proposed a Markov Chain Monte Carlo (MCMC) estimation approach based on Gibbs sampling updates for $\bm \beta, {\bm \mu}_g, \sigma_g, \lambda_g$ and a Metropolis-Hastings update for the latent positions ${\bf Z}_i$. This method has been implemented in the {\tt latentnet} \citep{latentnet} R package. The MCMC fitting routine is in function {\tt ergmm}. This function allows one to use certain model terms that do consider self-loops, an important characteristic of the ORIE network as discussed in section \ref{Sec:1}. We use the method described in \citet{latentnet} to setup the priors for all hyperparameters.

For the MCMC estimation, for each alternative model we used a warm-up period of 50,000 iterations,  then ran the MCMC routine for 5,000,000 more iterations, and finally collected statistics only every 100 iterations over 50,000 more iterations. 

\begin{figure}
\begin{center}
\resizebox{14cm}{10cm}{\includegraphics{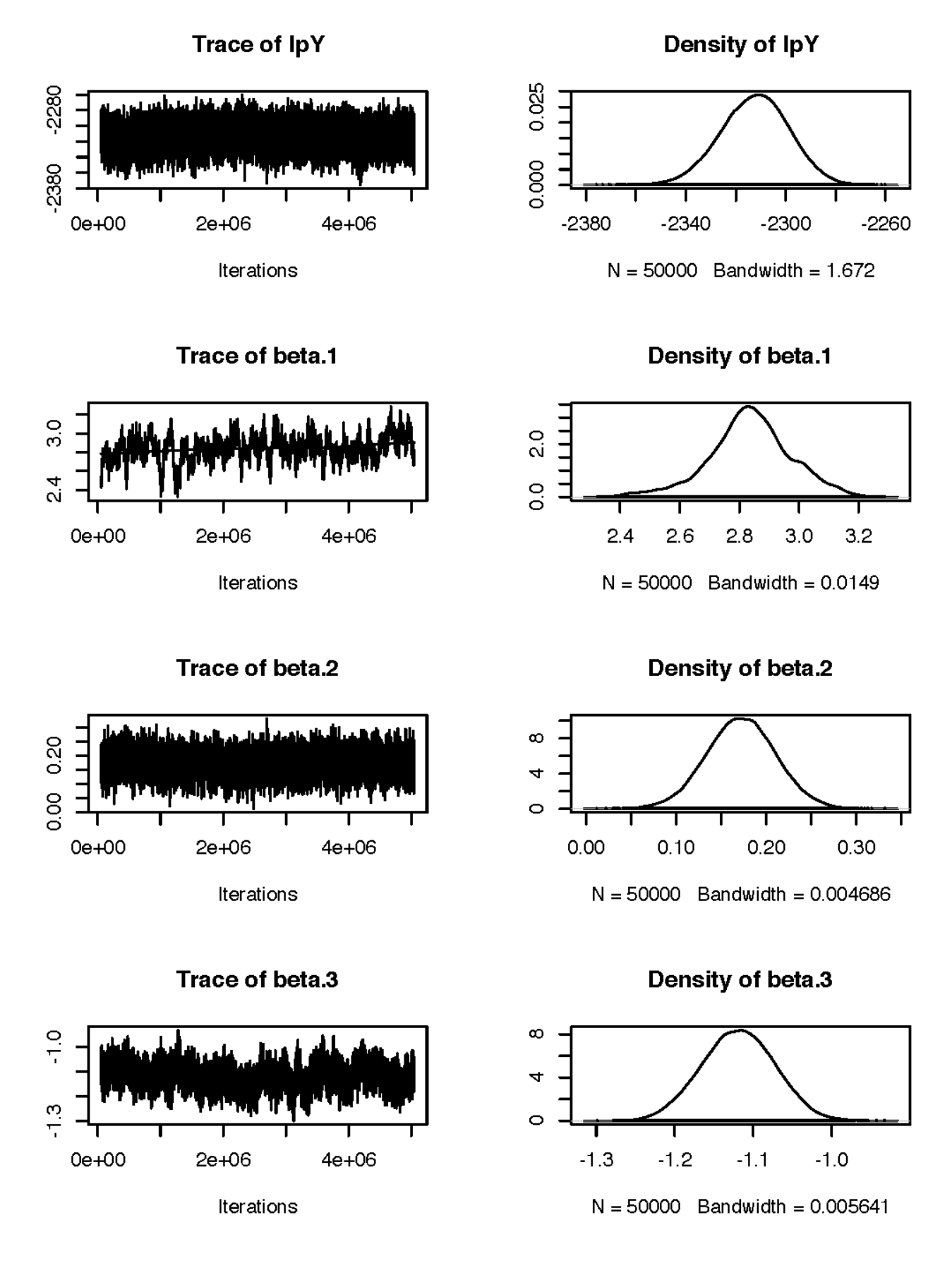}}
\caption{MCMC diagnostics for model (\ref{fitted}): log-likelihood and values of $\beta_0, \beta_1$, and $\beta_2$. All convergence diagnostics are adequate, i.e., trace plots are stable and therefore parameter values converge in distribution. Traces shown for last 5M iterations, distributions shown for last 50K iterations. } 
\label{fig:mcmc2}
\end{center}
\end{figure}

\begin{figure}
\begin{center}
\resizebox{16cm}{18cm}{\includegraphics{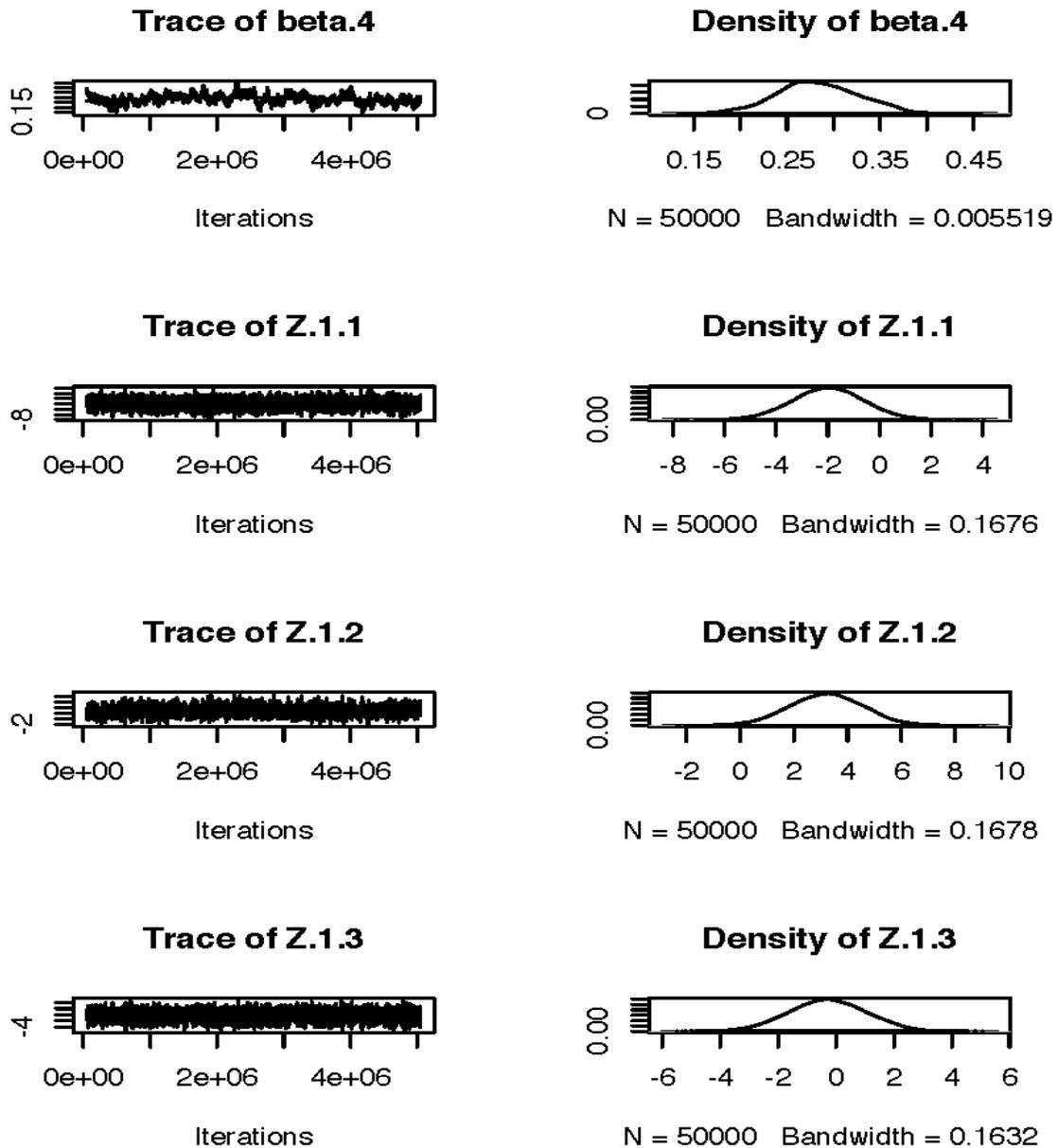}}
\caption{MCMC diagnostics for model (\ref{fitted}), (cont.): values of $\beta_4, Z_1, Z_2$, and $Z_3$. All convergence diagnostics are adequate, i.e., trace plots are stable and therefore parameter values converge in distribution. Traces shown for last 5M iterations, distributions shown for last 50K iterations.} 
\label{fig:mcmc3}
\end{center}
\end{figure}

There are not many techniques available for ERGM model selection. We selected a final model comparing the alternative models' Bayesian Information Criterion (BIC), considering whether the MCMC estimation procedure converged satisfactorily or not for a given model, and looking at model diagnostics involving the prediction performance of each model considered. Models with 2-dimensional ($d=2$) location vectors ${\bf Z}_i$ had difficulty converging; the chains did not show adequate mixing. For $d=3,G=2$ (2 groups or clusters in 3 dimensions), BIC=5171 but the $\beta_j$ coefficients failed to converge; for $d=3$ and $G=3$, BIC=5221 (somewhat worse) but it had satisfactory MCMC convergence for all its parameters (see Figures \ref{fig:mcmc2} and \ref{fig:mcmc3}).\\  

The fitted final model, obtained from the posterior means of all parameters, is:
\begin{eqnarray}
\log(\mu_{ij}) &=& \log(E[Y_{ij}|{\bm \beta, \bf x, Z}]) = \beta_0 + \beta_1 x_{ij} + \beta_2 x_i + \beta_3 x_j - |{\bf Z}_i - {\bf Z}_j|, \; i, j=1,...,n \nonumber\\
&=& \underbrace{2.8363}_{(2.52,3.10)} + \underbrace{0.1722}_{(0.09,0.24)} \; x_{ii}  \underbrace{- 1.1190}_{(-1.21,-1.02)} \; x_i + \underbrace{0.2828}_{(0.19,0.37)} \; x_j -  |{\bf Z}_i - {\bf Z}_j|
\label{fitted}
\end{eqnarray}
where $x_{ij} = \log(\mbox{MVS}_2(i))$ if $i=j$ and $x_{ii}=0$ for $i \neq j$ is a self ``loop" covariate, modeling self-hires, $x_i = \log(\mbox{MVS}_2(i))$ is a ``sender" covariate and $x_j=\log(\mbox{MVS}_2(j))$ is a ``receiver" covariate \citep{latentnet}. Numbers in the under braces are the 95\% credible posterior intervals for each coefficient, they all exclude zero. The sender and receiver covariates model the propensity of a department to either provide or receive faculty. The motivation for this model was that the MVS$_2$ indices can provide an indication of the number of faculty flowing between departments. Taking the log(MVS$_2$) implies we assume effects are linear, given this is a log-linear model. For instance, suppose MVS$_2(1)=1$ and MVS$_2(2)=2$. Then the fitted model predicts $\mu_{11}=16.94$ self hires on average for the top ORIE department.

\subsection{Model prediction performance}

\begin{figure}
\begin{center}
\resizebox{8cm}{8cm}{\includegraphics{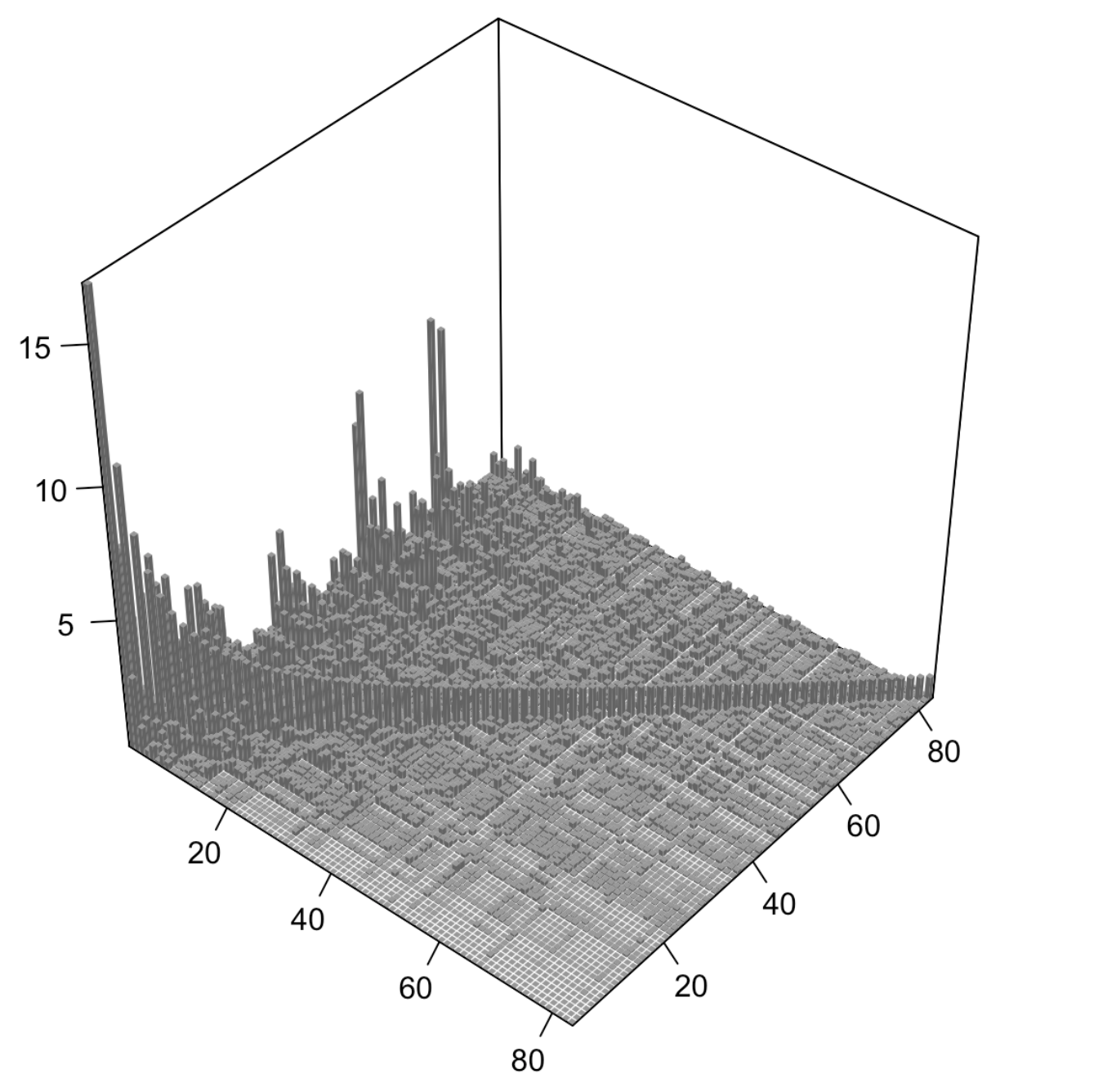}}
\caption{Predicted mean number of faculty hires in the ORIE network ($\hat \mu_{ij}$) given by the fitted latent location ERGM model (\ref{fitted}). Height equals expected number of faculty, axes are the 83 departments sorted according to MVS$_2$. Compare with the observed number of hires, the plot on the right of Figure \ref{fig:hillside}.}
\label{fig:Pred} 
\end{center}
\end{figure}

The fitted latent location ERGM model (\ref{fitted})  adequately predicts the expected value of the number of hires within and between departments. Using $i=j$ in (\ref{fitted}), the predicted mean self-hire values $\mu_{ii}$ are obtained, and these are contrasted with the observed self-hired values in the ORIE network in the top plot of Figure \ref{fig:0} (red line), showing excellent agreement. Figure \ref{fig:Pred} diagrammatically shows the fitted expected number of hires between all 83 ORIE departments, which can be compared to the plot on the right of Figure \ref{fig:hillside}, again demonstrating very close behavior to the observations. More specific model diagnostics for ERGM models are based on Monte Carlo simulations of the fitted network and comparison of the observed statistics to those in the ensemble of simulations.  

\begin{figure}[H]
\begin{center}
\begin{tabular}{ll}
\resizebox{8cm}{6cm}{\includegraphics{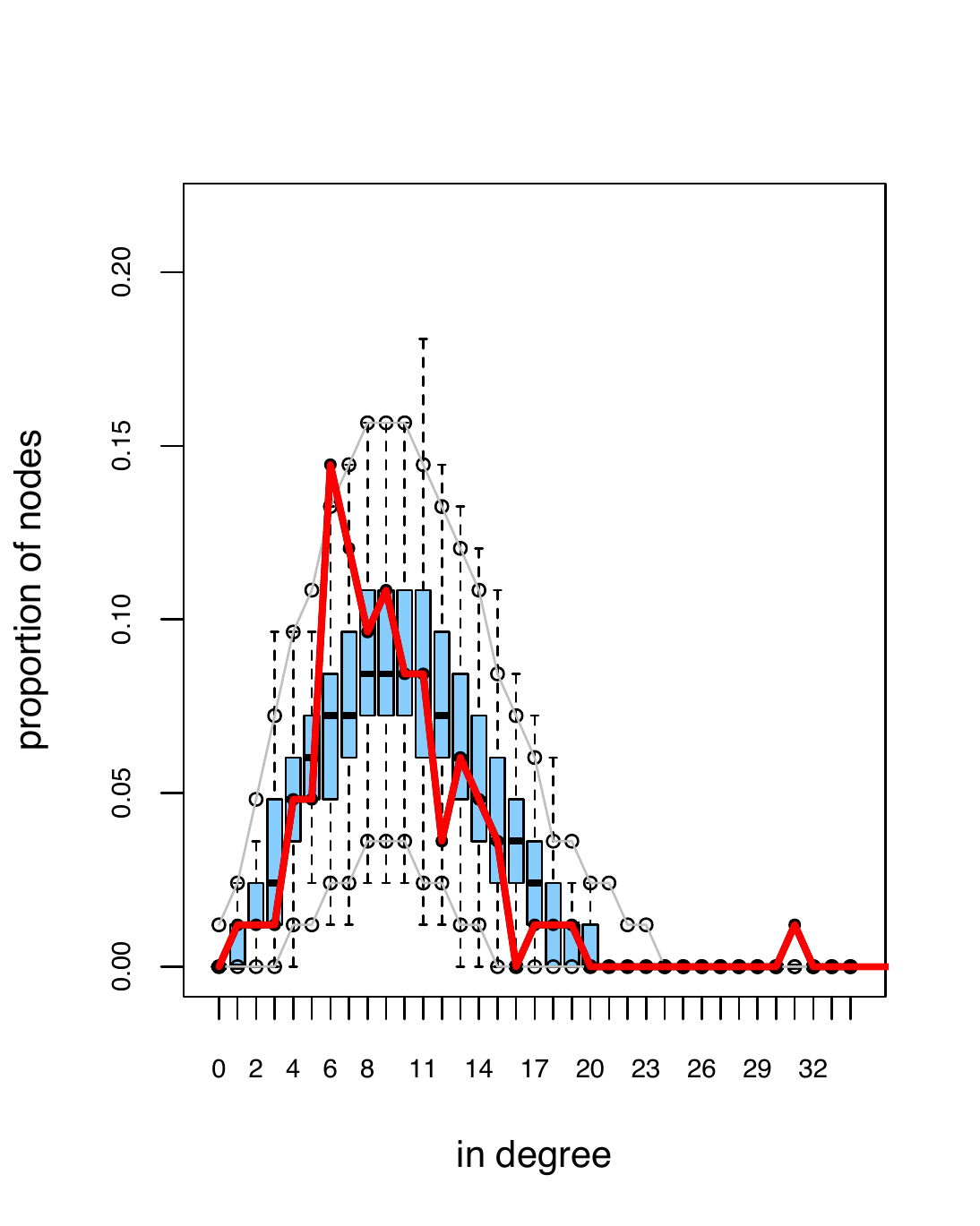}}&
\resizebox{8cm}{6cm}{\includegraphics{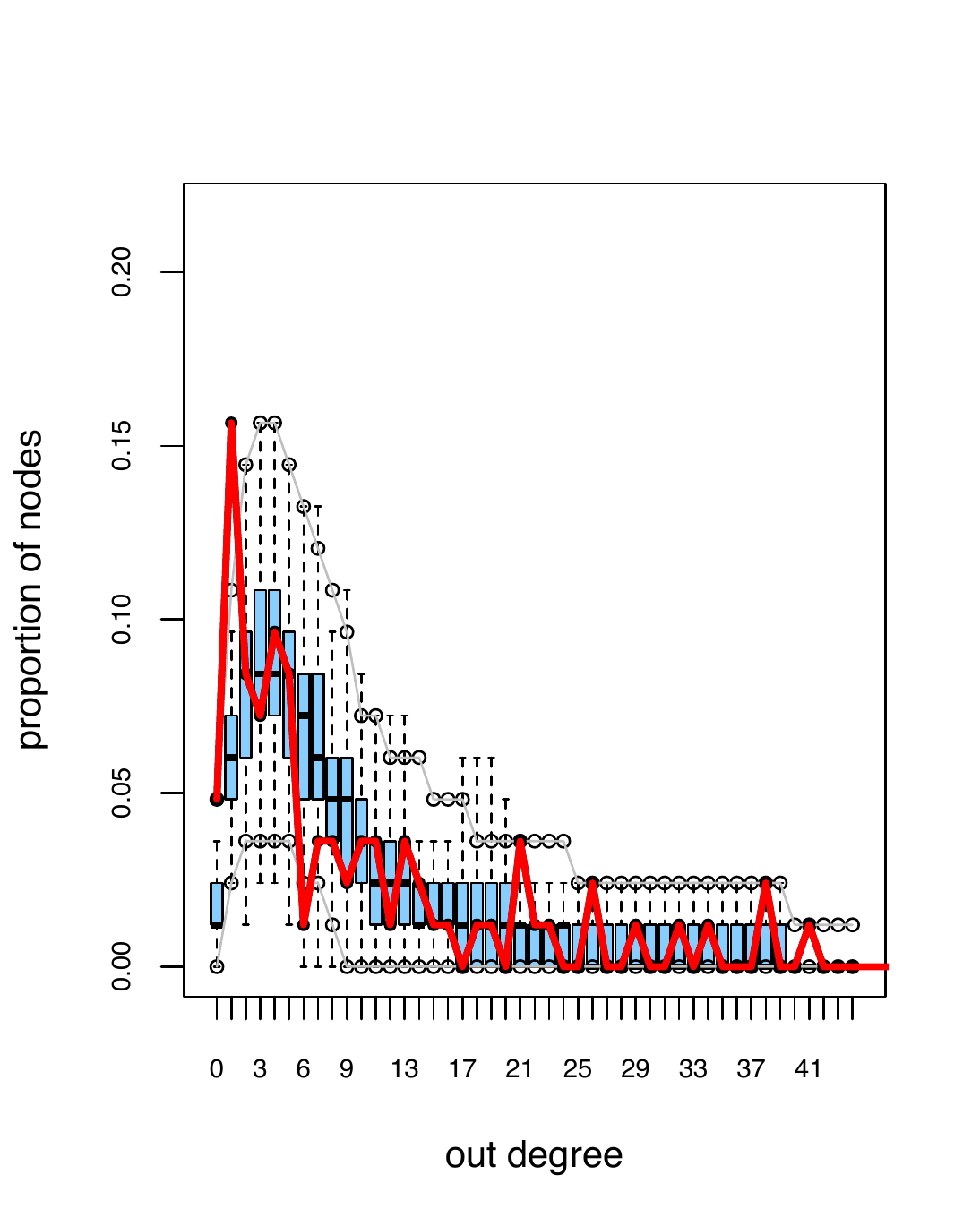}}
\end{tabular}
\caption{Diagnostic plots for the ERGM model (\ref{fitted}) based on 1000 simulations. Left: in-degree distribution. Right: out degree distribution. Boxplots give the interquartile ranges and the whiskers the extreme values simulated from the posterior of the fitted model (\ref{fitted}), red continuous lines are the observed values. With exception of a slight over-generation of vertices with in-degree 6 an 7 and out-degree equal to one, the fitted model reproduces the observed degree distributions.} 
\label{fig:Simulations0}
\end{center}
\end{figure}

\begin{figure}
\begin{center}
\begin{tabular}{ll}
\resizebox{8cm}{5.8cm}{\includegraphics{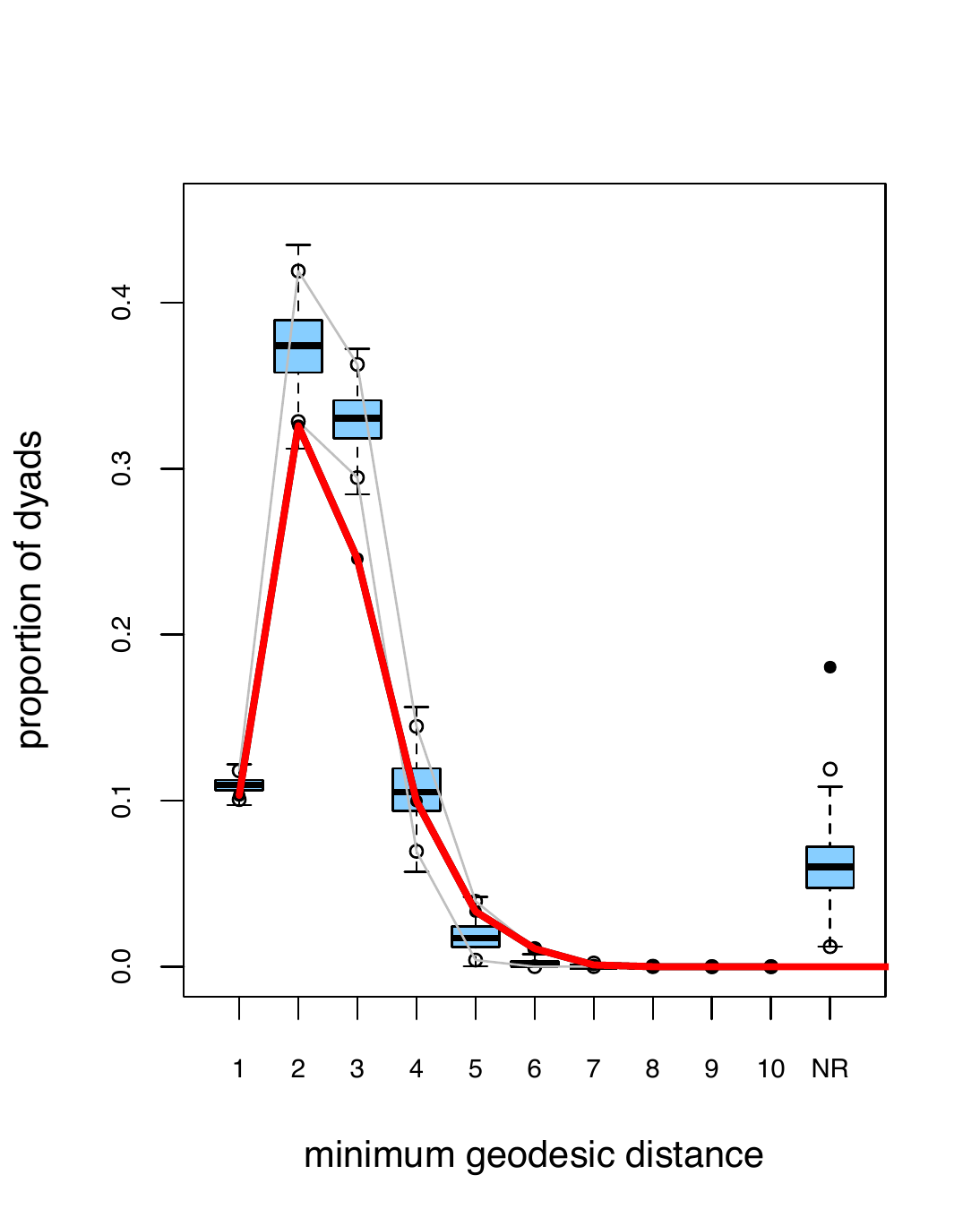}}&
\resizebox{8cm}{6cm}{\includegraphics{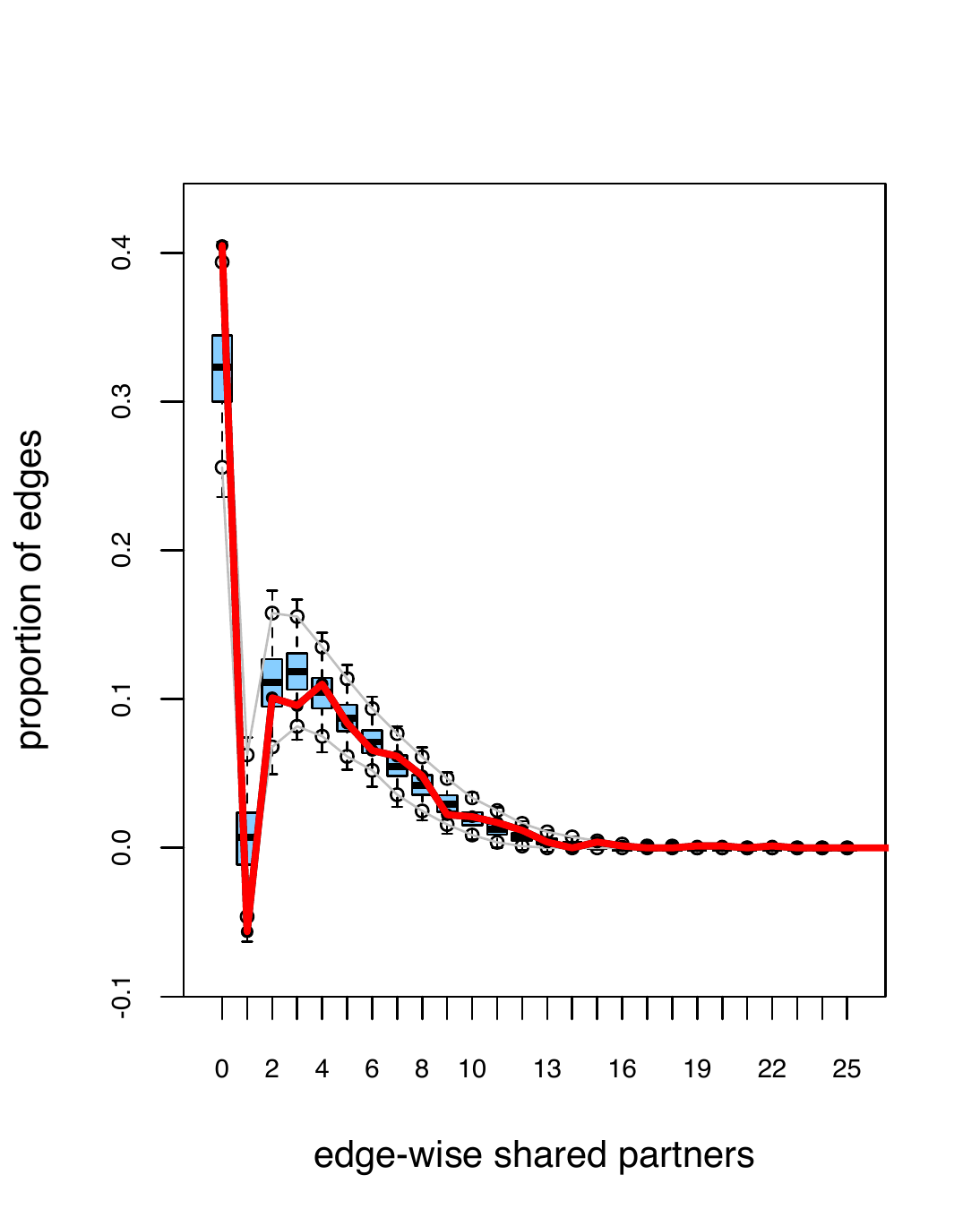}}
\end{tabular}
\caption{Diagnostic plots for the latent ERGM model (\ref{fitted}) based on 1000 simulations. Left: geodesic distance distribution. Right: edge-wise shared partner distribution. The number of edge-wise shared partners is defined as the number of edges $e_{ij}$ such that both $i$ and $j$ have $k$ common neighbors ("shared partners"). Boxplots give the interquartile range and the whiskers the extreme values simulated from the posterior of the fitted model (\ref{fitted}), red continuous lines are the observed values.  The fitted model is able to reproduce the observed statistics. Note how almost 40\% of the edges in the ORIE network have no common neighbors.} 
\label{fig:Simulations1}
\end{center}
\end{figure}

\begin{figure}
\begin{center}
\begin{tabular}{lll}
\resizebox{5cm}{5cm}{\includegraphics{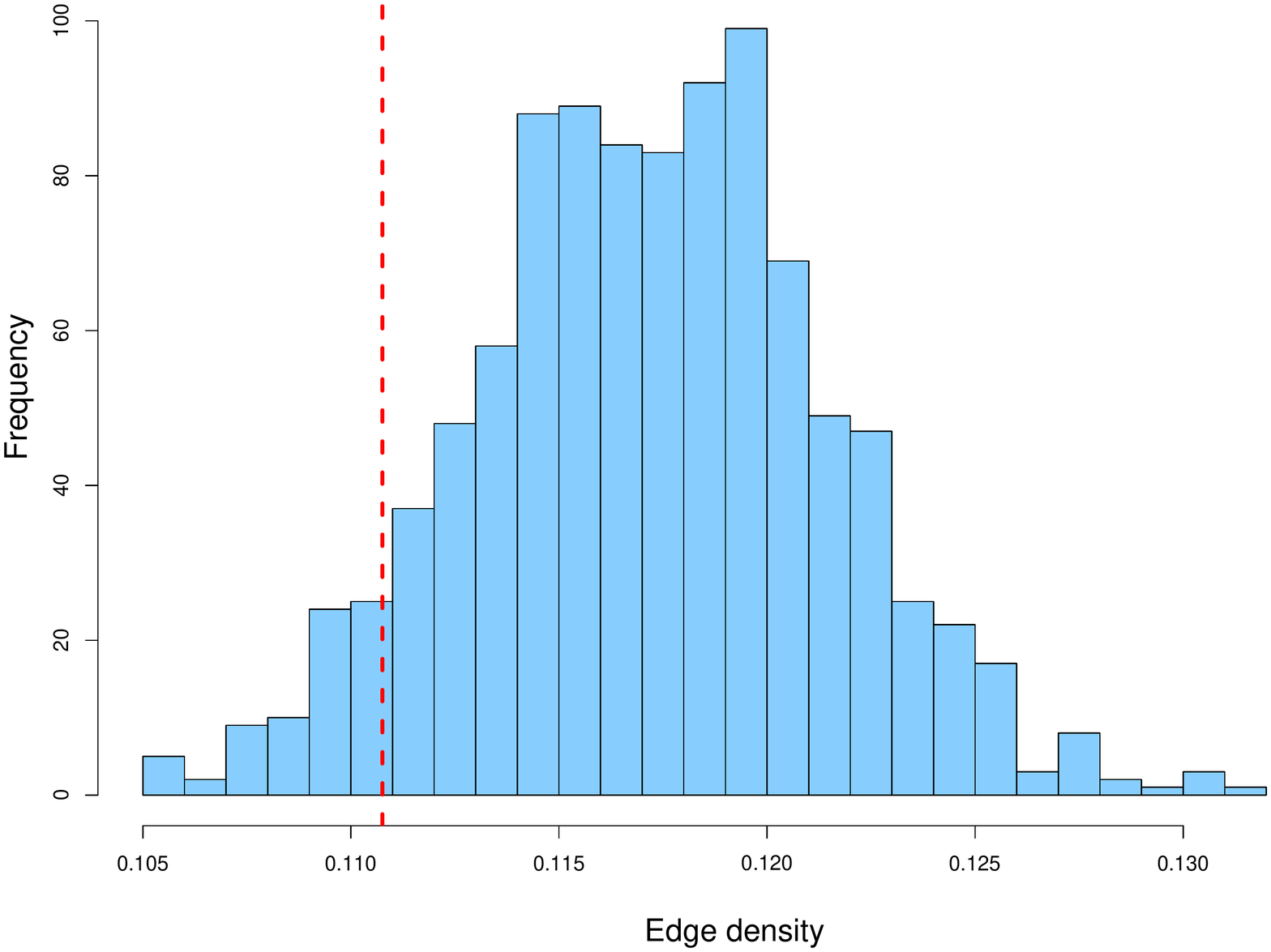}}&
\resizebox{5cm}{5cm}{\includegraphics{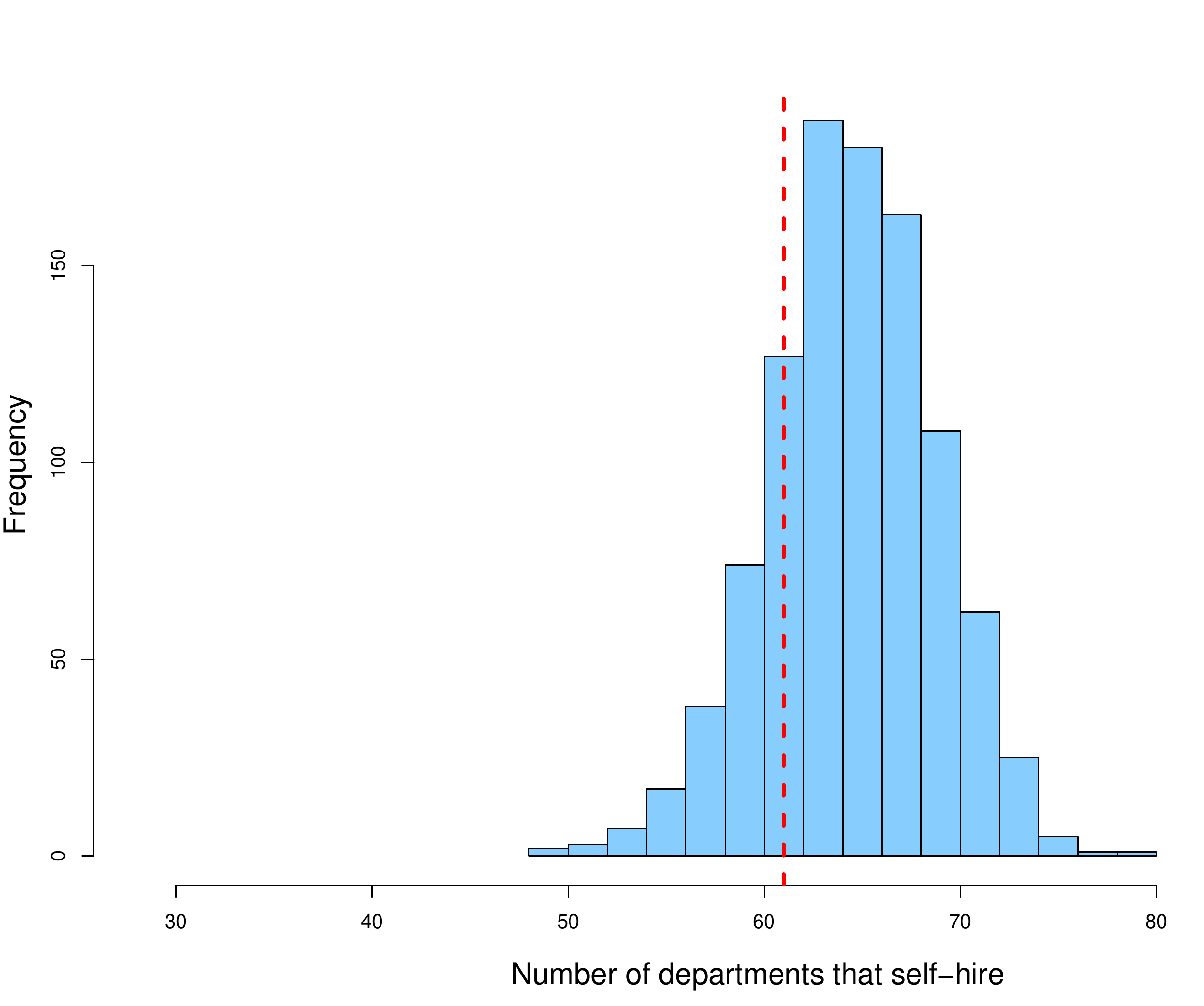}}&
\resizebox{5cm}{5cm}{\includegraphics{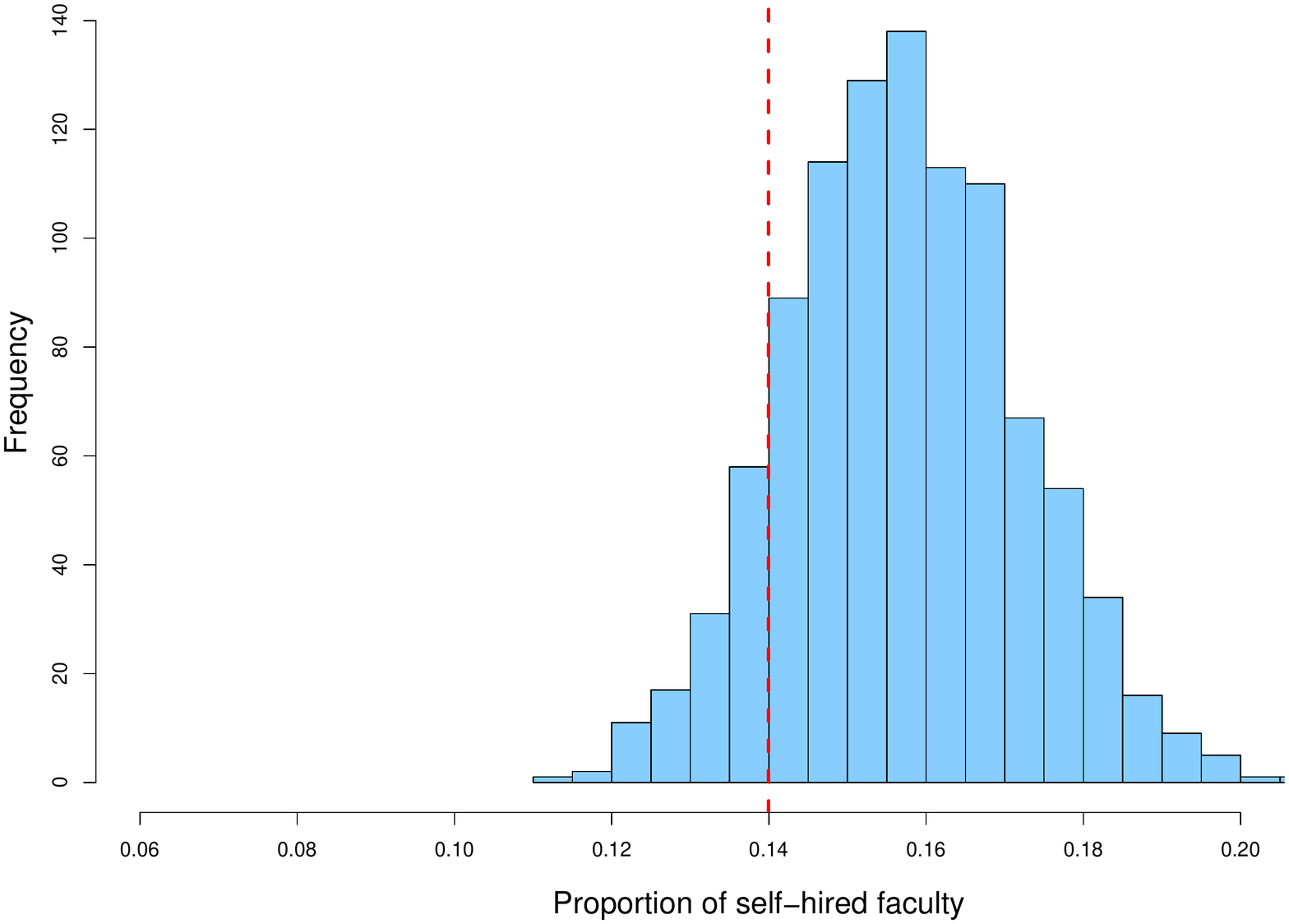}}
\end{tabular}
\caption{Diagnostic plots for the latent location ERGM model (\ref{fitted}) based on 1000 simulations. Left: edge density distribution. center: number of departments that self-hire. Right: proportion of self-hired faculty. Observed values in ORIE network are given by red vertical lines. In all cases, the fitted model reproduces the observed values in the actual network.} 
\label{fig:Simulations2}
\end{center}
\end{figure}

We performed posterior checks based on one thousand simulated networks from the posterior of the fitted model using the {\tt latentnet} R package. We compared posterior properties of the simulated networks against the corresponding statistics of the observed ORIE faculty network. Figure \ref{fig:Simulations0} contrasts the actual observed values for the in and out degree distribution (bold red) with the interquartile range of the simulated values from the posteriors  (blue boxplots). Both distributions can approximately generate the observations, including the very skewed out-degree distribution. There is some over-generation of nodes with in-degrees equal to 6 and 7, and over-generation of nodes with out-degrees equal to one edge. Figure \ref{fig:Simulations1} (left) contrasts posterior simulations of the minimum geodesic distances between the vertices and the actual values, which are reproduced well by the model. The plot on the right contrasts the posterior simulations vs. actual values for the edge-wise shared partners, defined as the number of edges $e_{ij}$ such that both $i$ and $j$ have $k$ common neighbors. The peculiar shape of this distribution is very well reproduced by the latent ERGM model. Finally, in Figure \ref{fig:Simulations2} (left) we first computed the posterior edge density distribution of the simulated networks and contrasted it with the actual value (red dotted line),  which falls among the simulated values, indicating the model can reproduce networks with the correct density of edges.  The number of departments that self-hire ranges from around 49 to 78 in the posterior simulated networks, and this includes the observed value (61) in the real ORIE network (red dotted line in Figure \ref{fig:Simulations2}, center). Finally, the proportion of Ph.Ds self hired by their original department  ranges from about 12\% to 20\%, again including the observed 14\% (red dotted line, Figure \ref{fig:Simulations2}, right).

We conclude from these ``goodness of fit" posterior simulations that the general structure of the ORIE faculty hired network is captured by the fitted latent ERGM (\ref{fitted}).

\subsection{Determination of groups of departments}

\begin{figure}
\begin{center}
\resizebox{11cm}{7cm}{\includegraphics{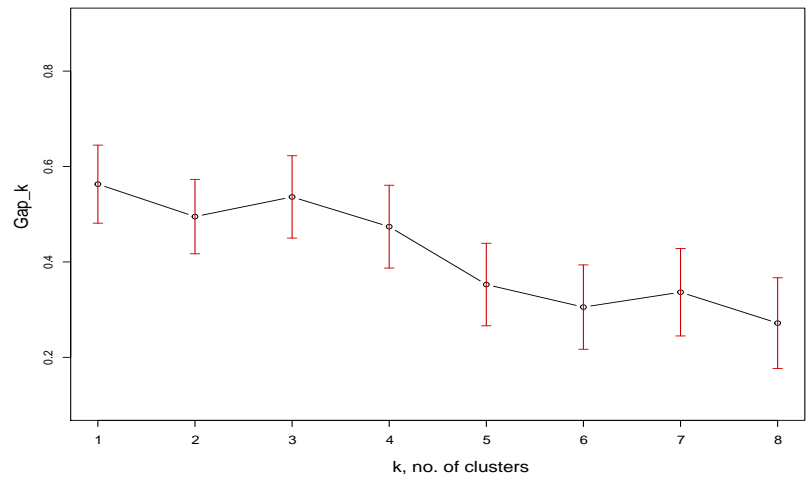}}
\caption{Simulated ``gap" statistics \citep{Gap} to determine the number of clusters in the latent variables ${\bf Z}_i$ identified in the exponential random graph model (\ref{fitted}). Either 1 or 3 clusters are identified, we chose 3 clusters due to better interpretability.} 
\label{fig:clusters}
\end{center}
\end{figure}

The latent ERGM model permits us to determine $G$ groups in which the nodes are clustered in the location space $\bf Z$. We found the clustering provided by the ERGM model unreliable, the clusters did not seem to form in any particular meaningful way. Instead,  to select groups of departments, we took the mean of the posterior of the latent locations ${\bf Z}_i$, $i=1,...,83$ and ran the PAM (Partitioned around medoids) algorithm  \citep{PAM} to find 3 groups (clusters) of departments. This provided a much better group separation than using the $\lambda_g$ proportions in (\ref{Zs}) as the basis of clustering. The number of clusters was found using PAM and the ``Gap" statistics \citep{Gap} that determines a goodness of clustering measure. The best number of clusters is either 1 (no clustering) or 3 from the gap statistics (see Figure \ref{fig:clusters}). We therefore form 3 groups or clusters of departments, and the groups are displayed in different color in Figure \ref{fig:LatentPositions} and listed in Appendix 1(Table \ref{tab:A1}).

\begin{figure}
\begin{center}
\begin{tabular}{ll}
\resizebox{8cm}{7cm}{\includegraphics{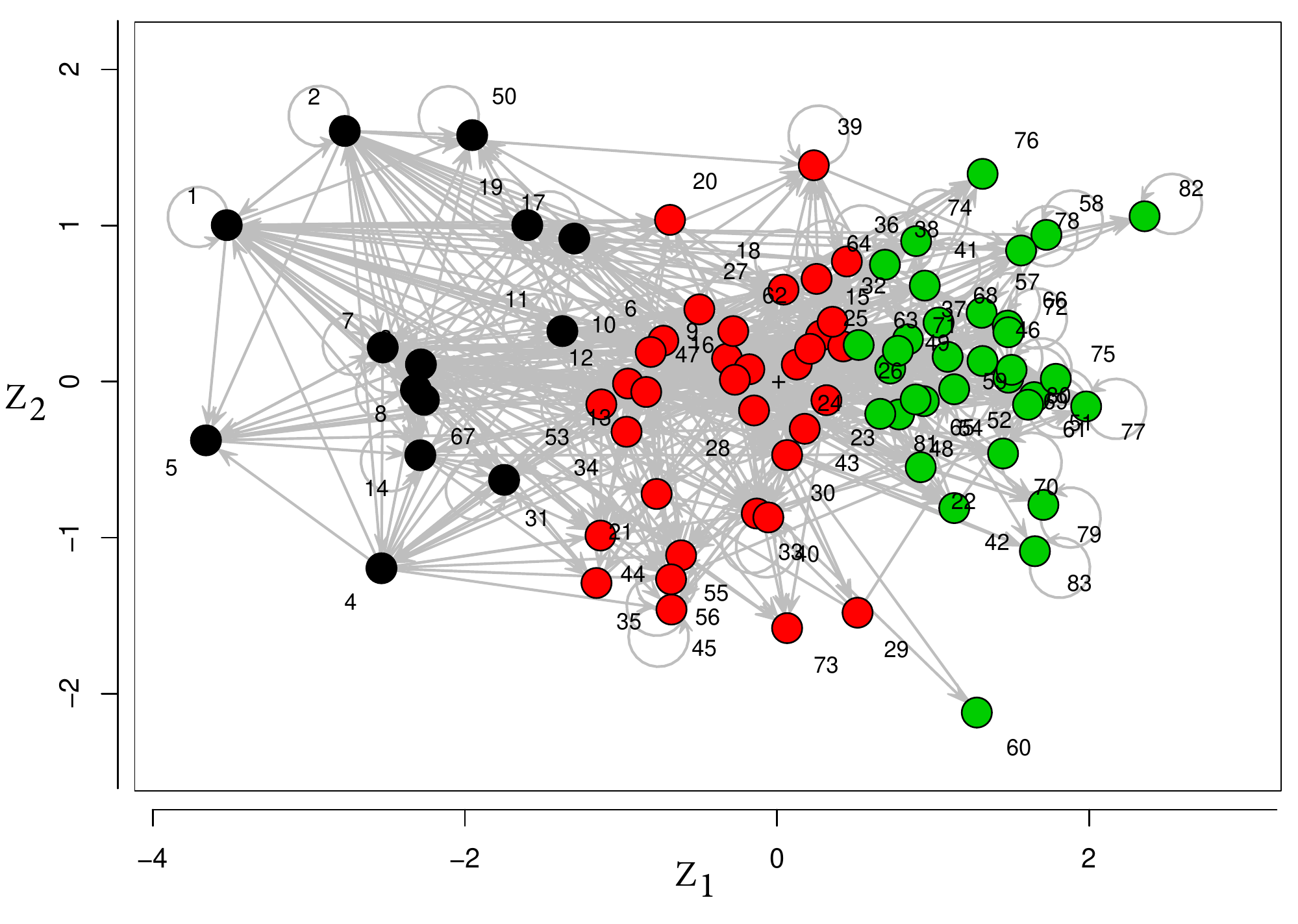}}&
\resizebox{8cm}{8cm}{\includegraphics{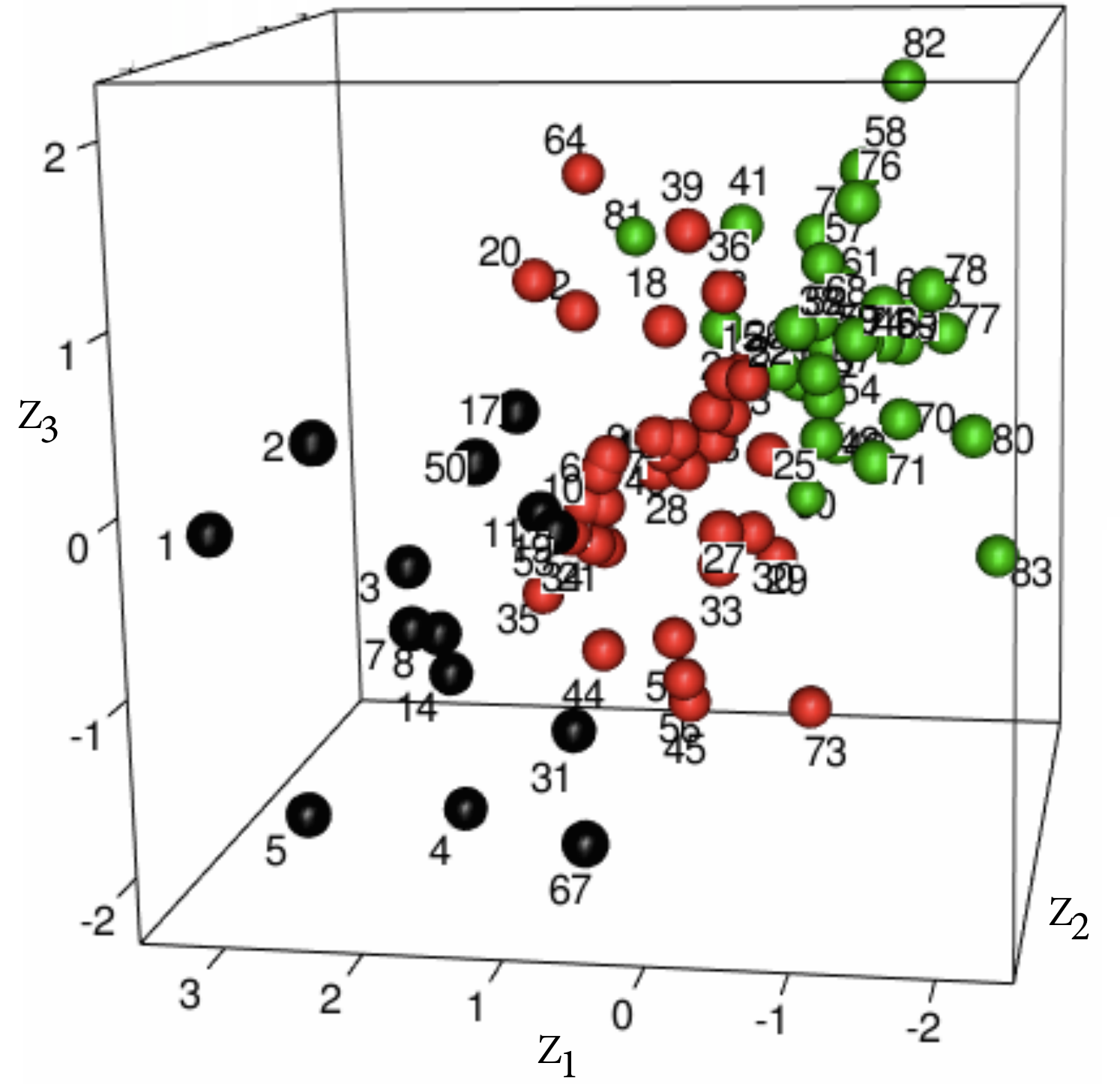}}
\end{tabular}
\caption{Latent Positions $Z_1$, $Z_2$, $Z_3$ after fitting (\ref{fitted}). Vertices are colored according to the three identified groups (black=group 1, red=group 2, and green = group 3). Left: plot over first two latent variables, Right: plot over the 3-dimensional latent space (edges not shown). Numbers correspond to MVS$_2$ indices, see Table \ref{tab:A1}. Evident features are the few connections between groups 1 and 3, and how group 2 contains departments with a high degree of ``betweenness" \citep{Newman}.} 
\label{fig:LatentPositions}
\end{center}
\end{figure}

\begin{figure}
\begin{center}
\resizebox{13cm}{11cm}{\includegraphics{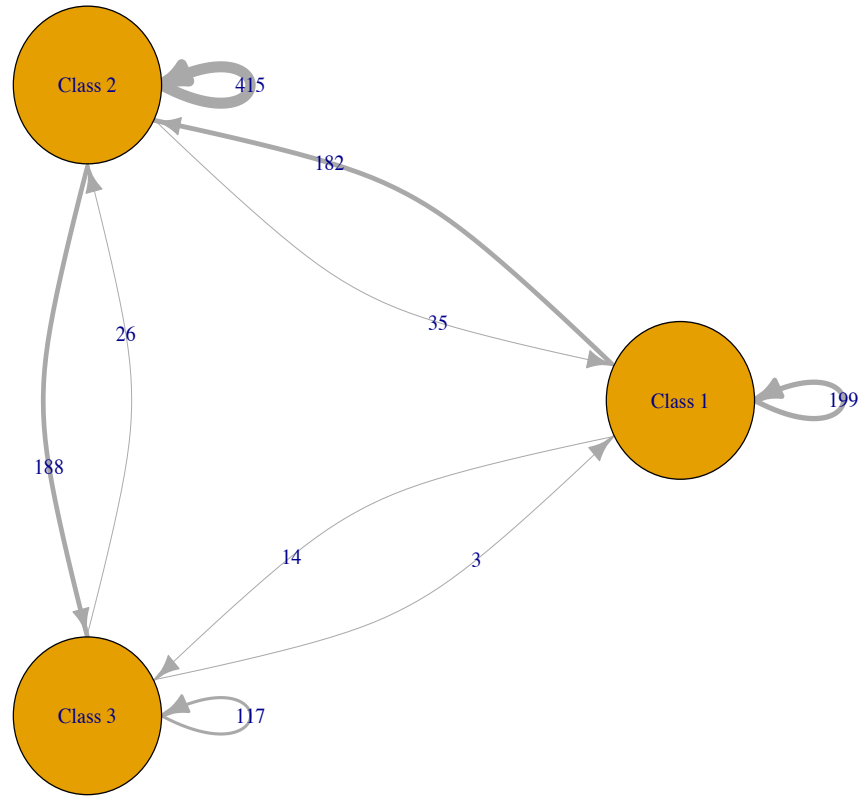}}
\caption{ Simplified ORIE network, showing the faculty hires between the three identified groups of departments. Edge width is proportional to the number of faculty hires, which are the numbers shown on each edge. The bulk of the hire-placements occurs within group. Note also how there are very few connections between groups 1 and 3, with departments in group 3 contributing only to three hires in departments in group 1. } 
\label{fig:SimplerNetwork}
\end{center}
\end{figure}

It is notable how well the first latent group includes only departments who more consistently hire from top departments in the hierarchical MVS$_2$ order. 
Using the three groups of departments, we can reduce the ORIE network to a simple aggregated network where faculty flow within and between the three  groups (Figure \ref{fig:SimplerNetwork}). Most of the hires (62 \%) take place within groups, only 38\% is between groups. Note how groups 3 and 1 are very thinly connected: only 3 faculty receiving their Ph.D. in the 34 departments in group 3 have been hired by the 14 departments in group 1. Inversely, only 14 Ph.D.'s from group 1 have been hired in departments from group 3. Group 3 has also provided few (26) faculty to group 2. This indicates that the ORIE network is strongly separated in clusters, with departments in the bottom of the hierarchy (group 3) producing almost no faculty for those on the top. 
Interestingly, the ``intermediate" departments in group 2 (35 departments) include the top 7 departments as ranked by ``betweenness" and overall have the lowest (i.e. most important) betweenness  average ranks (38.3 for group 2, 42.2 for group 1, 44.7 for group 3). It could be argued that the intermediate departments in group 2 keep the ORIE as a single, connected  discipline. An interesting question for further work is to determine if the degree of interaction between groups 1 and 3 is higher or lower than the interaction threshold between different (but close) disciplines, such as OR and Computer Science, for instance. 



\section{Discussion and Conclusions.}

We have demonstrated the existence of a near linear hierarchy in the ORIE faculty hiring network through the application of modern techniques in statistical analysis of social networks that exclude subjective assessments. 
We provided an approximate linear hierarchy index (Minimum Violation and Strength, MVS$_2$) obtained through optimizing bootstrapped networks sampled from the original ORIE network and therefore not sensitive to the unequal density of edges (and that stand in contrast to published rankings).
Single indices of hierarchy, however, do not capture all important features in a complex network, for instance, the highly connected core of the departments in the top of the hierarchy, the high incidence of self-hires, or the skewed out degree distribution.
This is the reason we propose and fit the latent location Exponential Random Graph Model (ERGM) of section \ref{Sec:5} and the associated cluster analysis as a better approach for  understanding the hiring patterns in the ORIE faculty network (and of any such hiring network), resulting in groups that do not necessarily follow the MVS$_2$ ranks. The model  successfully captures the main characteristics of the ORIE faculty network, including its node degree distributions, its incidence of self-hiring, and its edge density distribution among others. This model allowed us to simplify the ORIE network to one with only three groups of departments, with most hiring taking place within groups, also showing the little interaction between groups at the extremes of the hierarchy. To be in the first group, a department needs to consistently hire from top departments, even if it does not place that many Ph.D.'s in other top departments, a non-existent notion in competitive sports and ecology. Furthermore, departments in the intermediate group act as a ``link" between the other two groups, keeping the network connected as a single discipline, with departments in this group having among the highest ``betweenness" importance \citep{Newman}.\vspace{0.1cm}


 ORIE is a field with a high self-hiring rate across the whole network compared to Computer Science and Business Schools (close to two and three times, respectively). 
A high self-hiring rate may reduce the transfer of information and discipline-specific knowledge (including curriculum innovations) between departments. \vspace{0.1cm}

We are aware hiring decisions are not only made based on the ``prestige" of the potential ``sender" and ``receiver" departments. Besides the obvious impact of each candidate's  Ph.D. research credentials, they are also a consequence of complex politics, personalities and styles of the faculty in the ``receiver" department and its college administration. Sometimes the reputation of the Ph.D. adviser, and not only that of the sending department, matters in a hiring decision. Sometimes  department chairs overwrite the majority opinion of their own faculty in hiring decisions, thus some actual hires may not represent the normal behavior of a department. These aspects of the hiring-placement process can be thought to add ``noise" to the collected data, which is already sparse at lower levels of the hierarchy. 
In our analysis we handled this noise via permutation tests or bootstrapping when finding evidence of a hierarchy or trying to determine an index, or through a statistical model for the network. But as discussed in section \ref{zeroes}, zero valued edges do not necessarily indicate the absence of a dominance relation. Unfortunately, it is not possible to determine from the available data whether an ``observational zero" in a social network implies the lack of a dominance relation or rather its presence, with the dominated party trying to avoid interactions with the dominant one (i.e., departments {\em avoiding} hiring from higher ranked departments). \vspace{0.1cm}


The analysis presented in this paper refers to data collected at a single point in time (summer 2016). This neglects the {\em dynamic} effect of hires over the years which could be modeled if more complete data about the years of subsequent employments of each single professor in the network were available (our datasets contain partial data, too incomplete to attempt such analysis and this is left for future work). Dynamic exponential random graph models seem a good way to study such dynamic networks.

\bibliography{IEORSocialNetwork}
\bibliographystyle{authordate1}

\section*{Appendix 1: Institution data}

The list of ORIE departments was formed by merging those Ph.D. granting departments in the 2016 US News \& World report "Industrial/Manufacturing/Systems Engineering" graduate rankings (accessed 5/12/2016) with those in the 2011 National Research Council \citep{NRC} rankings for "Operations Research, Systems Engineering and Industrial Engineering". The NRC lists both departments and programs, so an institution may appear more than once; we included each institution only once in the network (their NRC ranking shown below corresponds to the highest ranking of either program or department) and collected the faculty information as if it were a single department. Only departments with existing web pages as of May 2016 and that have a Ph.D. program were included (this excluded U. of Nebraska-Lincoln). The NRC rankings listed in Table \ref{tab:A1}   below are those given by the average of the 5\% and 95\% ``R" rankings, listed in standard competition order. There are Industrial Engineering departments which are merged with other disciplines in a single department (e.g., Mechanical and Industrial Engineering, or Computer Systems and Industrial Engineering); an effort was made to include only those faculty in the ORIE field. It was assumed, in the absence of information, that a faculty working in an ORIE department obtained his/her Ph.D. in the ORIE department of the listed institution (a number of faculty only lists their alma mater institution but not their alma mater's department). Also, faculty that received their Ph.D. outside of the USA (or outside of our list of departments) were not considered. Only faculty in tenured or tenure-track positions were included, although it was not always clear if some faculty positions were tenured or not. 

It is important to point out that the edges ($v_i,v_j)$ of the ORIE network are formed by faculty {\em currently} (as of summer 2016) in department $j$ who received their Ph.D. in department $i$. This evidently neglects the movement of faculty through some of other intermediate departments  between $i$ and $j$ and is a potential source of error. Our dataset contains partial information with respect to these intermediate departments where each faculty worked before their current job, information too incomplete to attempt an analysis.

The final ORIE dataset is contained in 2 files, one for the 83 institutions (vertices) and its attributes, and one for the 1179 faculty (edges) and its attributes. The data collection effort was undertaken in summer 2016 over a 4 week span. After data gathering, all faculty data was checked manually for errors by 3 persons. Table \ref{tab:A1}  contains the attributes in the institutions file, which includes ranks of various measures of vertex importance and the latent groups found in section \ref{Sec:5}.

\setcounter{table}{0}
\renewcommand{\thetable}{A\arabic{table}}

\begin{scriptsize}
\begin{table}
  \centering
  \caption{Latent group membership and importance measure ranks, ORIE departments, sorted by group and then by MVS$_2$ index.}
     \begin{tabular}{l|l|llllllllllll}
    {}&{}&\multicolumn{11}{|c}{Ranks}\\
    \hline
    {Institution}	&	{Group}	&	MVS$_2$	&	{MVS$_1$}	&	MVR	&	USNews	&	{NRC}	&	{In-deg}	&	{Out-deg}	&	{Eigen.}	&	{PageR.}	&	{Bet}	&	{Hub}	&	{Auth.} \\
    \hline
     Stanford  	&	1	&	1	&	2	&	2	&	4	&	1	&	5	&	2	&	2	&	2	&	17	&	2	&	2	\\
      UC Berkeley 	&	1	&	2	&	1	&	1	&	2	&	4	&	36	&	6	&	3	&	3	&	26	&	3	&	18	\\
    MIT 	&	1	&	3	&	3	&	3	&	6	&	3	&	3	&	1	&	1	&	1	&	18	&	1	&	1	\\
    Carnegie M. 	&	1	&	4	&	4	&	4	&	 \textit{NA} 	&	7	&	79	&	12	&	6	&	8	&	14	&	8	&	59	\\
    Princeton  	&	1	&	5	&	5	&	5	&	 \textit{NA} 	&	18	&	66	&	33	&	8	&	16	&	68	&	15	&	16	\\
    Cornell  	&	1	&	7	&	7	&	7	&	7	&	8	&	16	&	10	&	4	&	6	&	25	&	7	&	12	\\
    Columbia  	&	1	&	8	&	8	&	8	&	11	&	20	&	22	&	17	&	16	&	12	&	58	&	17	&	6	\\
    Northwest.   	&	1	&	11	&	11	&	11	&	4	&	6	&	20	&	15	&	10	&	9	&	39	&	10	&	28	\\
    U. Penn 	&	1	&	14	&	15	&	18	&	28	&	14	&	14	&	24	&	7	&	17	&	43	&	20	&	7	\\
    USC 	&	1	&	17	&	19	&	19	&	12	&	29	&	12	&	28	&	22	&	22	&	35	&	25	&	10	\\
    UT (Aus) 	&	1	&	19	&	18	&	16	&	19	&	23	&	46	&	29	&	12	&	15	&	33	&	22	&	21	\\
    UNC (Ch-H) 	&	1	&	31	&	35	&	42	&	 \textit{NA} 	&	34	&	32	&	58	&	23	&	45	&	48	&	38	&	36	\\
    Naval Post. 	&	1	&	50	&	83	&	75	&	23	&	 \textit{NA} 	&	23	&	69	&	69	&	78	&	70	&	60	&	14	\\
    Wash. U. 	&	1	&	67	&	66	&	43	&	39	&	 \textit{NA} 	&	69	&	71	&	71	&	68	&	67	&	71	&	34	\\
     \hline
    U. Michigan 	&	2	&	6	&	6	&	6	&	2	&	5	&	11	&	3	&	9	&	5	&	5	&	4	&	8	\\
    Purdue   	&	2	&	9	&	9	&	9	&	9	&	9	&	9	&	5	&	13	&	7	&	6	&	6	&	13	\\
    GA Tech 	&	2	&	10	&	10	&	4	&	1	&	2	&	5	&	4	&	1	&	5	&	1	&	5	&	3	\\
    U. Illi. (UC) 	&	2	&	12	&	14	&	14	&	15	&	33	&	13	&	13	&	17	&	14	&	28	&	14	&	5	\\
    U. Wis (Ma) 	&	2	&	13	&	12	&	12	&	7	&	9	&	21	&	14	&	15	&	11	&	10	&	9	&	22	\\
    U. Florida 	&	2	&	15	&	13	&	13	&	19	&	23	&	56	&	9	&	19	&	13	&	36	&	16	&	38	\\
    Penn St. 	&	2	&	16	&	16	&	15	&	12	&	12	&	7	&	8	&	18	&	10	&	4	&	13	&	15	\\
    Ohio St.  	&	2	&	18	&	21	&	21	&	17	&	9	&	19	&	11	&	25	&	19	&	12	&	19	&	19	\\
    U. Minn. 	&	2	&	20	&	20	&	20	&	32	&	46	&	61	&	39	&	37	&	38	&	61	&	29	&	31	\\
    U. Maryland 	&	2	&	21	&	17	&	17	&	 \textit{NA} 	&	16	&	60	&	22	&	11	&	21	&	37	&	11	&	32	\\
    U. Pitt 	&	2	&	23	&	24	&	24	&	23	&	39	&	25	&	16	&	32	&	24	&	19	&	24	&	20	\\
    VPI 	&	2	&	24	&	25	&	25	&	9	&	15	&	10	&	7	&	26	&	18	&	7	&	12	&	25	\\
    U.  Arizona 	&	2	&	25	&	27	&	27	&	28	&	35	&	53	&	26	&	41	&	37	&	55	&	27	&	51	\\
    NC State 	&	2	&	26	&	33	&	32	&	12	&	12	&	6	&	19	&	40	&	30	&	15	&	30	&	29	\\
    Lehigh 	&	2	&	27	&	23	&	26	&	18	&	28	&	39	&	34	&	30	&	29	&	47	&	43	&	40	\\
    SUNY Buf 	&	2	&	28	&	29	&	31	&	28	&	37	&	28	&	21	&	27	&	25	&	20	&	26	&	33	\\
    U. Miss (Co) 	&	2	&	29	&	26	&	23	&	58	&	31	&	78	&	54	&	34	&	51	&	69	&	45	&	63	\\
    Rutgers 	&	2	&	30	&	28	&	30	&	21	&	38	&	47	&	35	&	21	&	23	&	34	&	44	&	49	\\
Texas A\& M	&	2	&	32	&	31	&	34	&	15	&	24	&	15	&	20	&	20	&	20	&	2	&	23	&	30	\\
    U. Virginia 	&	2	&	33	&	37	&	35	&	28	&	23	&	27	&	37	&	48	&	42	&	38	&	34	&	23	\\
    U. Mass. 	&	2	&	34	&	39	&	46	&	36	&	50	&	8	&	25	&	29	&	31	&	23	&	18	&	9	\\
    Boston U. 	&	2	&	35	&	30	&	28	&	39	&	30	&	62	&	60	&	47	&	57	&	66	&	53	&	24	\\
    U. Arkansas 	&	2	&	36	&	40	&	41	&	39	&	57	&	31	&	44	&	39	&	48	&	40	&	49	&	43	\\
            RPI 	&	2	&	39	&	41	&	37	&	21	&	19	&	37	&	36	&	55	&	41	&	46	&	39	&	45	\\
    U. Wash. 	&	2	&	40	&	42	&	40	&	26	&	35	&	57	&	48	&	49	&	52	&	59	&	42	&	35	\\
    Iowa St. 	&	2	&	43	&	45	&	47	&	26	&	32	&	26	&	40	&	35	&	39	&	21	&	41	&	37	\\
    U. Conn. 	&	2	&	44	&	59	&	54	&	 \textit{NA} 	&	54	&	35	&	57	&	58	&	67	&	63	&	31	&	39	\\
    G. Wash. U. 	&	2	&	45	&	44	&	51	&	53	&	53	&	43	&	42	&	33	&	40	&	50	&	37	&	27	\\
    Arizona St. 	&	2	&	47	&	48	&	49	&	23	&	21	&	1	&	18	&	28	&	27	&	3	&	21	&	11	\\
    Northeastern 	&	2	&	53	&	81	&	82	&	36	&	27	&	4	&	70	&	70	&	79	&	71	&	51	&	4	\\
    Stevens 	&	2	&	55	&	61	&	74	&	39	&	 \textit{NA} 	&	17	&	49	&	50	&	55	&	53	&	40	&	17	\\
     G. Mason 	&	2	&	56	&	51	&	60	&	32	&	 \textit{NA} 	&	30	&	38	&	36	&	43	&	22	&	35	&	42	\\
    NJIT 	&	2	&	62	&	79	&	80	&	 \textit{NA} 	&	57	&	34	&	83	&	82	&	83	&	82	&	80	&	26	\\
    Case West. 	&	2	&	64	&	32	&	53	&	38	&	43	&	77	&	59	&	14	&	26	&	41	&	36	&	55	\\
    Worcester P 	&	2	&	73	&	74	&	64	&	53	&	 \textit{NA} 	&	73	&	80	&	75	&	80	&	74	&	82	&	46	\\
   \end{tabular}%
  \label{tab:A1}%
\end{table}
\end{scriptsize}

\newpage
\setcounter{table}{0}
\begin{scriptsize}
 \begin{table}[htbp]
  \centering
  \caption{Latent group membership and importance measure ranks, ORIE departments, sorted by group and then by MVS$_2$ index (cont.).}
    \begin{tabular}{l|l|llllllllllll}
    {}&{}&\multicolumn{11}{|c}{Ranks}\\
    \hline
{Institution}	&	{Group}	&	MVS$_2$	&	{MVS$_1$}	&	MVR	&	USNews	&	{NRC}	&	{In-deg}	&	{Out-deg}	&	{Eigen.}	&	{PageR.}	&	{Bet}	&	{Hub}	&	{Auth.} \\
    \hline
    U.  Iowa 	&	3	&	22	&	22	&	22	&	39	&	22	&	80	&	31	&	24	&	35	&	54	&	28	&	71	\\
    U. S. Florida 	&	3	&	37	&	38	&	36	&	46	&	54	&	54	&	27	&	42	&	34	&	11	&	33	&	67	\\
    Oklahoma St 	&	3	&	38	&	34	&	38	&	39	&	41	&	40	&	30	&	38	&	36	&	45	&	32	&	44	\\
    Kansas St. 	&	3	&	41	&	50	&	39	&	46	&	47	&	49	&	46	&	62	&	49	&	65	&	57	&	48	\\
         U. Illi. (Chi) 	&	3	&	42	&	36	&	33	&	46	&	45	&	71	&	52	&	45	&	53	&	62	&	48	&	61	\\
    Texas Tech  	&	3	&	46	&	47	&	48	&	53	&	40	&	55	&	23	&	51	&	33	&	9	&	47	&	68	\\
    U. Oklahoma 	&	3	&	48	&	53	&	56	&	46	&	50	&	42	&	53	&	57	&	61	&	49	&	58	&	50	\\
    Clemson  	&	3	&	49	&	58	&	58	&	32	&	47	&	29	&	51	&	63	&	58	&	44	&	61	&	52	\\
    Miss. U. 	&	3	&	51	&	43	&	44	&	58	&	62	&	50	&	32	&	44	&	32	&	31	&	50	&	78	\\
    Auburn  	&	3	&	52	&	49	&	52	&	32	&	42	&	48	&	41	&	54	&	47	&	29	&	52	&	66	\\
    Wayne St. 	&	3	&	54	&	54	&	61	&	53	&	44	&	44	&	64	&	52	&	60	&	42	&	62	&	41	\\
    U. Louisville 	&	3	&	57	&	57	&	50	&	66	&	62	&	51	&	55	&	65	&	54	&	56	&	65	&	58	\\
    U. Alabama 	&	3	&	58	&	52	&	45	&	 \textit{NA} 	&	61	&	67	&	47	&	64	&	44	&	57	&	63	&	57	\\
    UC Florida 	&	3	&	59	&	55	&	57	&	39	&	54	&	38	&	45	&	56	&	50	&	24	&	59	&	62	\\
    UT (Dal) 	&	3	&	60	&	46	&	29	&	58	&	 \textit{NA} 	&	83	&	74	&	53	&	77	&	76	&	55	&	82	\\
    U. Houston 	&	3	&	61	&	62	&	55	&	53	&	65	&	72	&	50	&	60	&	56	&	52	&	56	&	77	\\
    Air Force IT 	&	3	&	63	&	71	&	77	&	46	&	 \textit{NA} 	&	18	&	62	&	72	&	76	&	51	&	54	&	47	\\
    SUNY (Bin) 	&	3	&	65	&	80	&	81	&	58	&	57	&	24	&	65	&	76	&	75	&	75	&	67	&	53	\\
    Oregon St. 	&	3	&	66	&	77	&	72	&	46	&	49	&	41	&	63	&	73	&	69	&	72	&	69	&	69	\\
    Wichita St. 	&	3	&	68	&	69	&	78	&	66	&	 \textit{NA} 	&	33	&	76	&	67	&	73	&	60	&	78	&	56	\\
    W. Virginia  	&	3	&	69	&	63	&	70	&	66	&	 \textit{NA} 	&	52	&	56	&	59	&	59	&	13	&	66	&	64	\\
    UT (Arling) 	&	3	&	70	&	60	&	66	&	58	&	 \textit{NA} 	&	45	&	43	&	43	&	46	&	8	&	46	&	70	\\
    U. Tenn. 	&	3	&	71	&	56	&	69	&	58	&	62	&	58	&	61	&	31	&	28	&	27	&	68	&	72	\\
NCA\&T St	&	3	&	72	&	78	&	71	&	66	&	 \textit{NA} 	&	64	&	82	&	78	&	82	&	78	&	81	&	81	\\
    Florida St. 	&	3	&	74	&	64	&	79	&	66	&	 \textit{NA} 	&	59	&	66	&	46	&	62	&	16	&	64	&	65	\\
    U. NC (Ch.) 	&	3	&	75	&	65	&	73	&	58	&	 \textit{NA} 	&	63	&	73	&	66	&	74	&	32	&	73	&	76	\\
    U. Wis (Mil) 	&	3	&	76	&	76	&	65	&	58	&	 \textit{NA} 	&	74	&	81	&	77	&	81	&	77	&	83	&	60	\\
    Old Dom U  	&	3	&	77	&	67	&	76	&	 \textit{NA} 	&	66	&	65	&	68	&	68	&	66	&	64	&	74	&	83	\\
    Ohio U. 	&	3	&	78	&	68	&	83	&	66	&	 \textit{NA} 	&	68	&	67	&	61	&	63	&	30	&	72	&	80	\\
    New Mex St 	&	3	&	79	&	73	&	63	&	 \textit{NA} 	&	60	&	76	&	79	&	83	&	72	&	83	&	75	&	73	\\
    U.  Miami 	&	3	&	80	&	75	&	68	&	66	&	52	&	75	&	75	&	79	&	71	&	79	&	77	&	79	\\
    U. Ark. LR 	&	3	&	81	&	82	&	67	&	46	&	 \textit{NA} 	&	70	&	72	&	74	&	70	&	73	&	70	&	54	\\
    Montana St.  	&	3	&	82	&	72	&	59	&	 \textit{NA} 	&	 \textit{NA} 	&	82	&	78	&	81	&	65	&	81	&	76	&	74	\\
    Florida IT 	&	3	&	83	&	70	&	62	&	 \textit{NA} 	&	 \textit{NA} 	&	81	&	77	&	80	&	64	&	80	&	79	&	81	\\ 
\end{tabular}
\end{table}
\normalsize
\baselineskip=19pt   
\section*{Appendix 2: Statistical tests on the steepness of a linear hierarchy in a network}

\Citet{deVries2006} introduces the concept of the {\em steepness} of a linear hierarchy, namely, the size of absolute differences between adjacently ranked individuals in their overall success in winning dominance encounters. When these differences are large, the hierarchy is said to be steep and is called shallow otherwise. DeVries' steepness measure is based on the $D_i$ score by \citet{David}, a measure of the dominance success of an individual $i$ given by the unweighted and weighted sum of the individual's dyadic proportions of ``wins" (in our case, Ph.D. placed in other departments)  combined with an unweighted and weighted sum of its dyadic proportion of ``losses" (Ph.D's hired from other departments). The proportions of ``wins" $P_{ij}$ are defined as $P_{ij} = y_{ij}/n_{ij}$ where $y_{ij}$ is the $(i,j)$ entry in the sociomatrix $\bf Y$ and $n_{ij}$ is the total number of interactions between $i$ and $j$ (i.e., $n_{ij} = y_{ij}+y_{ji}$). David's (normalized) scores are defined as $
D_i = \left(w_i + w2_i - l_i - l2_i + n(n-1)/\right)/n
$
where $w_i = \sum_{j \neq i} P_{ij}$, $w2_i = \sum_{j \neq i} w_j P_{ij}$, $l_i = \sum_{j \neq i} P_{ji}$ and $l2_i = \sum_{j \neq i} l_j P_{ji}$. The factors containing the total number of individuals scale this term to make it vary between 0 and $n-1$. To measure the steepness of a hierarchy, order the individuals according to $D_i$. Call $R$ the rank of the individuals. Then a simple linear regression model $D = \beta_1 R + \beta_0$ is fit to the data. The estimated coefficient $\hat \beta_1$ is an estimate of the steepness of the hierarchy. 

To test for the significance of the steepness of a hierarchy using a permutation test, assume as null hypothesis that the steepness is that given by a randomly formed network. Fitting linear regression models $D = \beta_1 R + \beta_0$ from the randomly generated networks each with ranks $R$ will result in an empirical distribution of the $\hat \beta_1$ coefficient, which can then be compared to  the $\hat \beta_1$ estimate from the actual network. Small empirical p-values imply the hierarchy is significantly steeper than that given by a simple random graph.

Figure \ref{Fig:3} shows the regression models fitted to the David statistics for the ORIE, CS and Business networks. The ORIE and CS networks appear to have a steep hierarchy only for approximately the first 10 departments. This is in contrast to Business schools which have a steeper slope for about a third of the departments.  Figure \ref{Fig:4} indicates, however, that the slope is significantly different to that of a random graph in all 3 cases. The steepness results shown are related to the connectedness (density) of the 3 networks, since the CS network is the most sparse, followed by ORIE and then Business. 

\begin{figure}
\begin{center}
\begin{tabular}{ccc}
\resizebox{5cm}{4cm}{\includegraphics{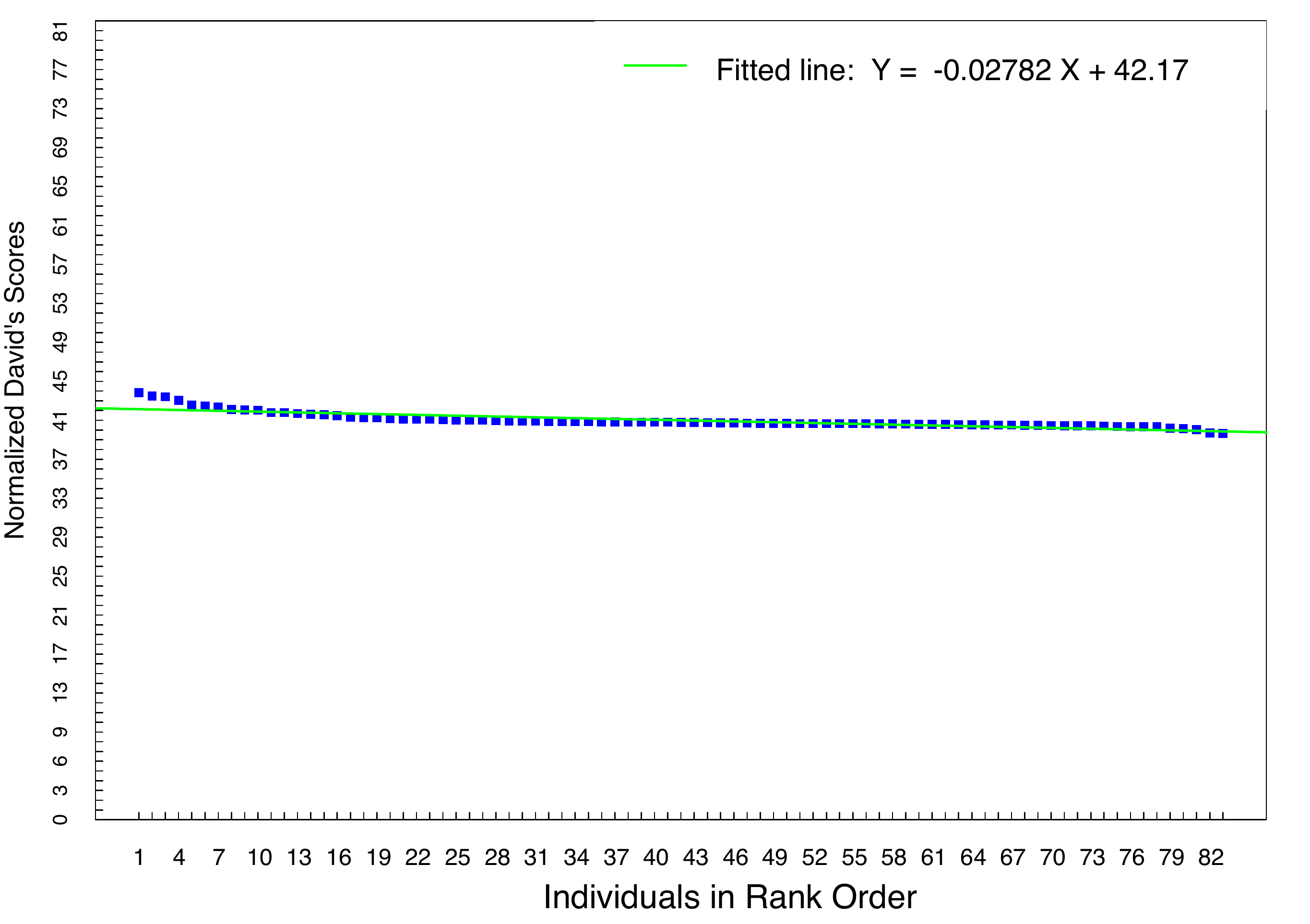}}&
\resizebox{5cm}{4cm}{\includegraphics{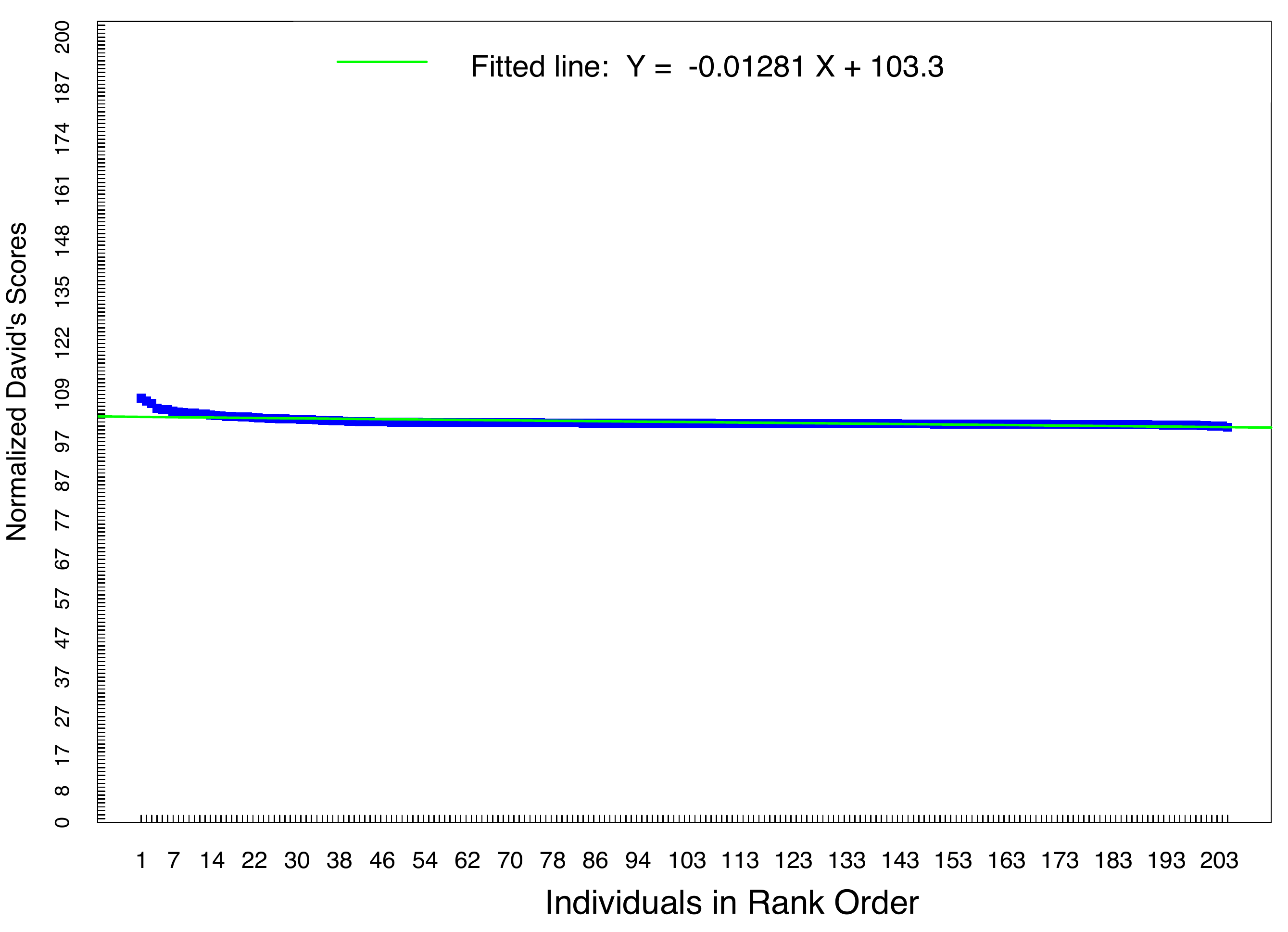}}&
\resizebox{5cm}{4cm}{\includegraphics{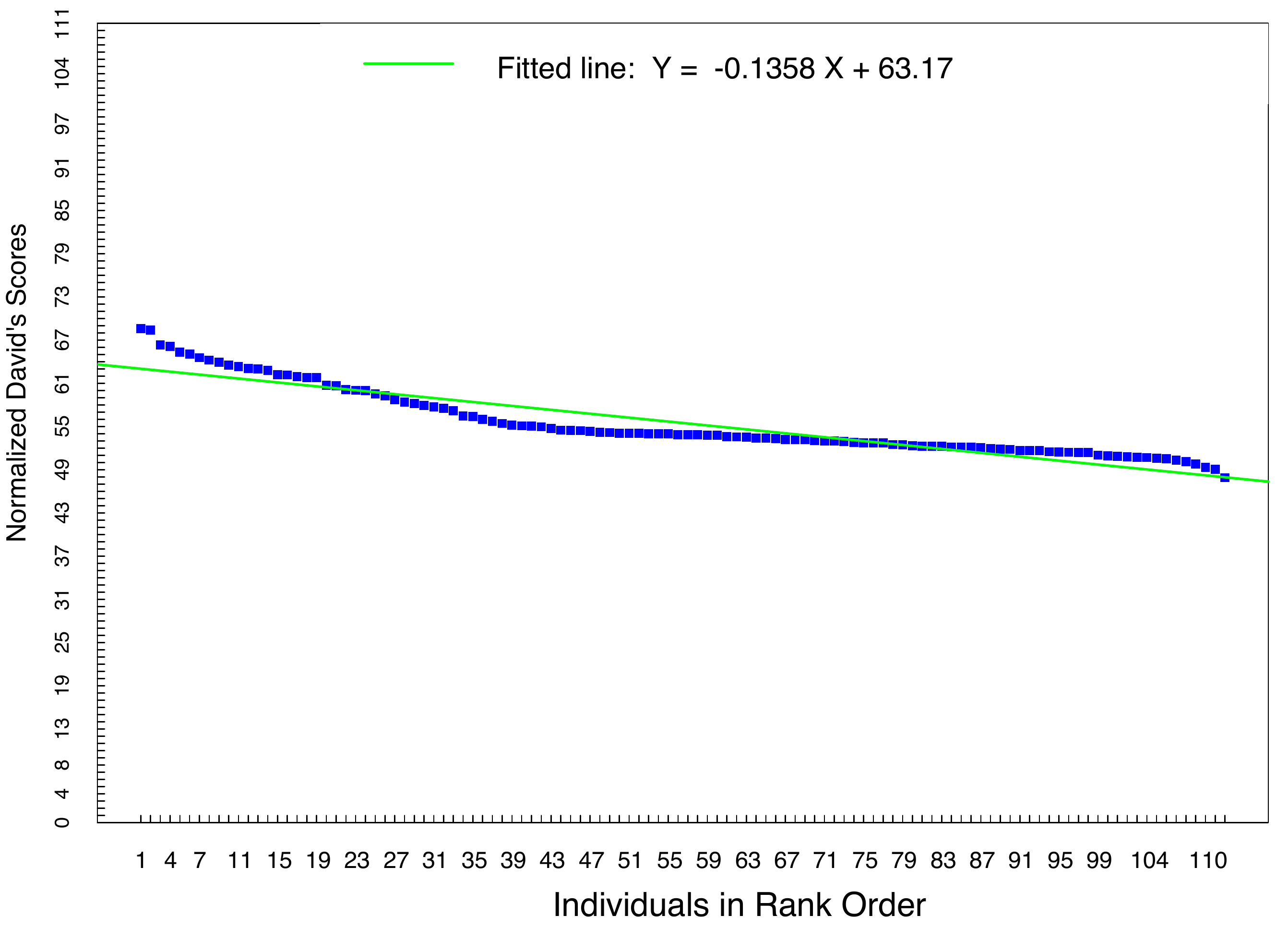}}\\
\end{tabular}
\caption{Fitted line and observed David's statistics for a linear hierarchy. Left: ORIE network ($n=83$, slope = -0.02782). Middle: Computer Science ($n=205$, slope = -0.01281). Right: Business ($n=112$, slope=-0.1358). Note how for ORIE and CS the observed absolute slope increases for the first 10 departments, indicating a steeper dominance at the top of the hierarchy relative to other departments, but otherwise the hierarchy is quite flat. } 
\label{Fig:3}
\end{center}
\end{figure}

\begin{figure}
\begin{center}
\begin{tabular}{ccc}
\resizebox{5cm}{4cm}{\includegraphics{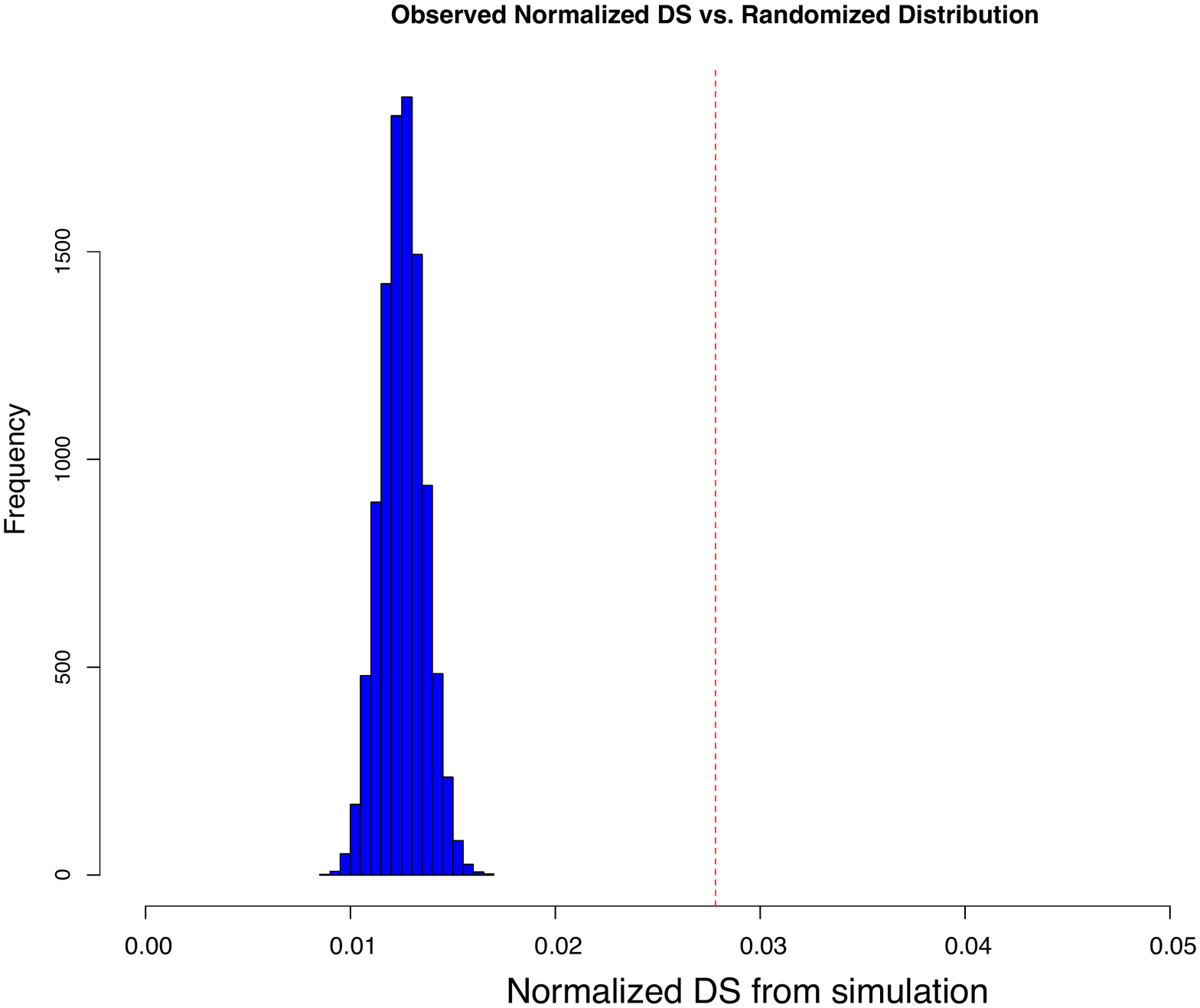}}&
\resizebox{5cm}{4cm}{\includegraphics{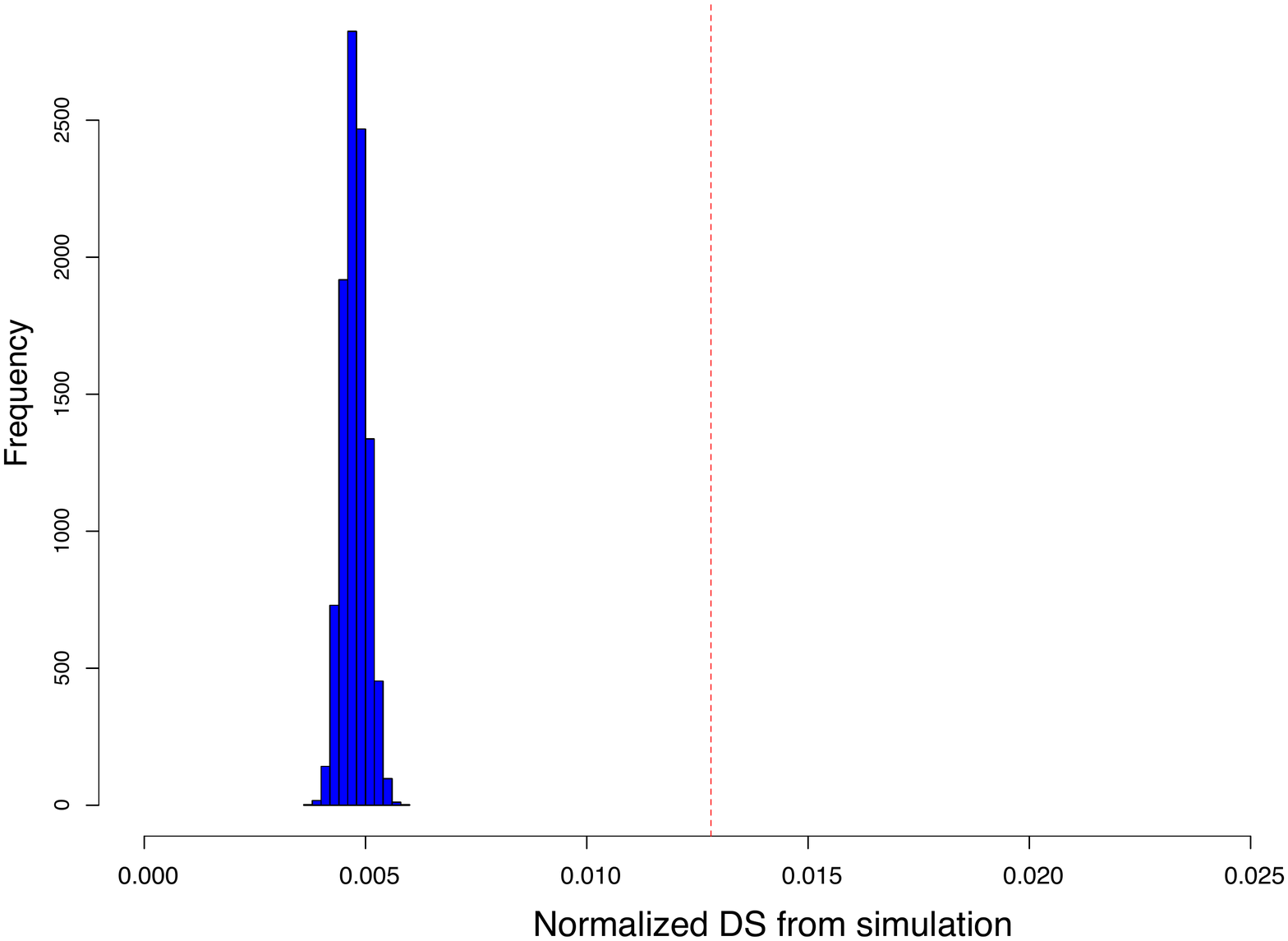}}&
\resizebox{5cm}{4cm}{\includegraphics{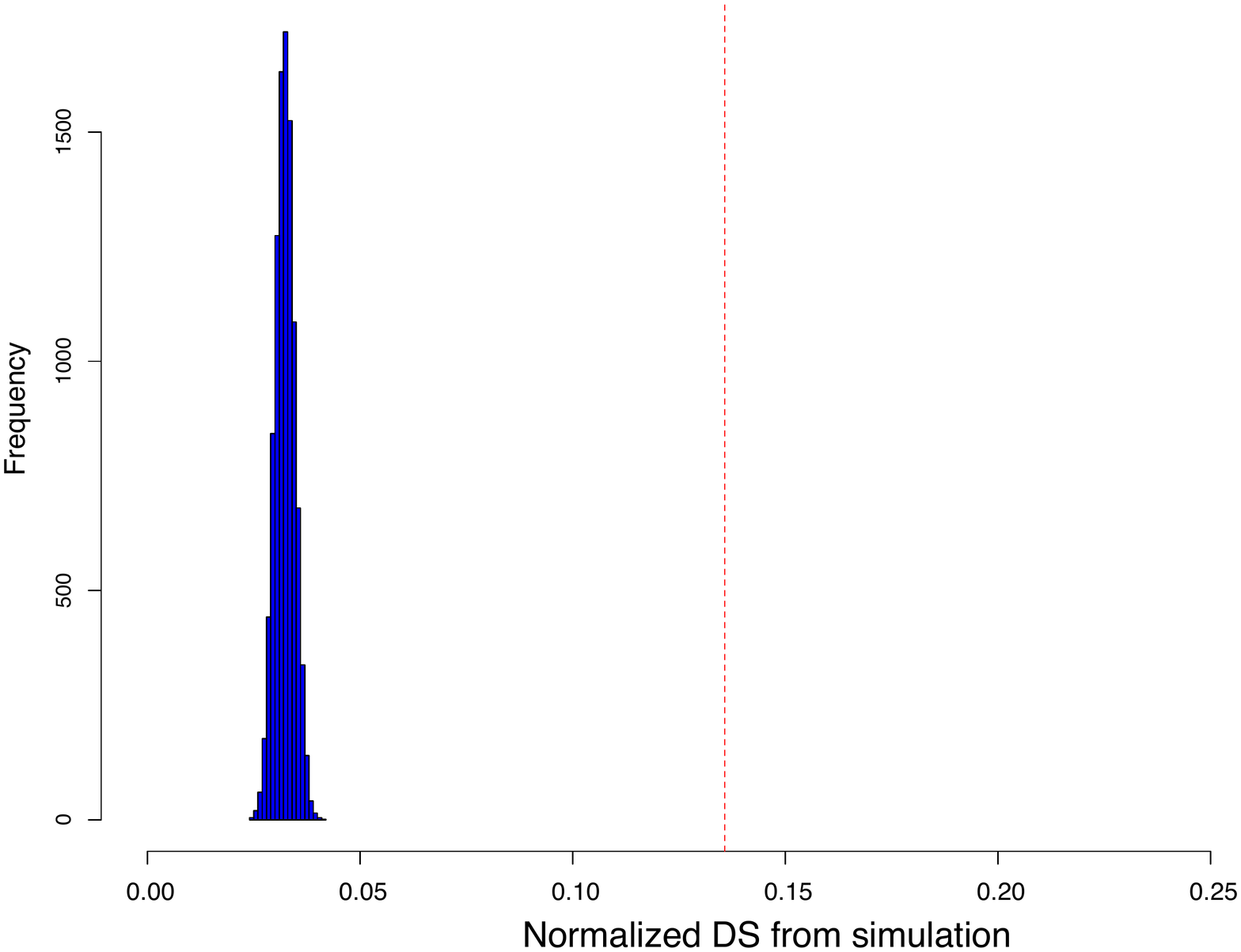}}\\
\end{tabular}
\caption{Randomized distribution (10K simulations) of the De Vries' test statistic for the significance of the steepness in the dominance hierarchy of (Left) the complete ORIE network, (Middle) the Computer Science network, and (Right) the Business network. Red vertical lines are the observed David's $D_i$ statistics. In all cases, the empirical p-values are 0.0, indicating significance steepness in the hierarchy of each faculty network. } 
\label{Fig:4}
\end{center}
\end{figure}

\section*{Appendix 3: Stochastic search algorithm for finding MVS rankings}

\makeatletter
\def\BState{\State\hskip-\ALG@thistlm}
\makeatother

\begin{algorithm}[H]
\caption{Minimum Violation and Strength (MVS) ranking of boostrapped networks}
\begin{algorithmic}[1]
\Procedure{optimize}{${\bf Y}, B$, burnin, iterations, interval}
\For {$b=1$ to $B$}
	\State ${\bf Y}_b \gets $ bootstrap($\bf Y$)
	\State $\pi_0({\bf Y}_b) \gets$ out degree rankings
	\For {$k=1$ to burnin}
		\State $\pi({\bf Y}_b) \gets $ SWAP(${\bf Y}_b, \pi_0({\bf Y}_b)$)
	\EndFor
	\For {$k=1$ to iterations}
		{\State $\pi_b({\bf Y}_b) \gets $ SWAP(${\bf Y}_b, \pi_0({\bf Y}_b)$)
		\If{ int$(k/\mbox{interval})-k/\mbox{interval}==0$} Save $\pi_b({\bf Y}_b)		$
		\EndIf}
	\EndFor
	\State MVS$_2({\bf Y}_b) \gets $ average(saved $\pi_b({\bf Y}_b)$'s)
\EndFor
\Return MVS$_2 \gets $ average(MVS$_2({\bf Y}_1),....,$MVS$_2({\bf Y}_B))$
\EndProcedure
\end{algorithmic}
\end{algorithm}

\begin{algorithm}[H]
\caption{Stochastic swapping to improve a given ranking}
\begin{algorithmic}[1]
\Procedure{swap}{$\bf Y, \pi(\bf Y)$}
\State Randomly select vertices $i$ and $j$ and swap them to form $\pi_{\mbox{\tiny new}}(\bf Y)$.
\If {$S(\pi_{\mbox{\tiny new}}({\bf Y})) > S(\pi({\bf Y}))$ or
 ($S(\pi_{\mbox{\tiny new}}({\bf Y})) = S(\pi(\bf Y))$ and
$S(\pi_{\mbox{\tiny new}}({\bf Y}))  < S(\pi(\bf Y))$)\\
$\quad$}
\State Accept the swap and \Return $\pi_{\mbox{\tiny new}}(\bf Y)$.
\Else $\quad$ \Return  $\pi(\bf Y))$
\EndIf
\EndProcedure
\end{algorithmic}
\end{algorithm}
\normalsize
\baselineskip=19pt
Algorithm 1 is a modification of the optimization approach in \cite{Clauset}, who found minimum violation rankings, adapted for finding MVS rankings. They reported exchanges of more than 2 vertices did not improve the solutions found with swapping pairs of vertices. The algorithm takes as initial ranking that of the out-degrees of each node gives preference to departments that place faculty in other departments. In analogy with Markov Chain Monte Carlo methods, the algorithm was run for a burn-in period of $10^5$ iterations (that are discarded), after which 1000 ranks $\pi_b(\bf Y)$ were saved every 100 iterations (i.e., iterations=$10^5$ and interval=100 in Algorithm 1). The averages of these 1000 ranks gives the MVS$_1$ or MVS$_2$ ranks. In addition, to incorporate the uncertainties  related to the low density areas of the ORIE network, which could be considered as ``noise", the optimization was repeated for $B=1000$ different bootstrapped networks, where the edges in each network were randomly sampled with replacement from the edges of the real ORIE network, with sampling probabilities proportional to the edge attributes $Y_{ij}$. The reported MVS$_1$ and MVS$_2$ ranks in Table \ref{tab:A1} correspond to the ensemble mean of the optimal ranks obtained from each of these bootstrapped replications, using either $S_{\mbox{\tiny MVS$_1$}}(\pi(\bf Y))$ or $S_{\mbox{\tiny MVS$_2$}}(\pi(\bf Y))$ in the SWAP procedure (Algorithm 2).

\end{scriptsize}
\end{document}